%% file: main.tex
\documentclass[twocolumn,twocolappendix]{aastex631}

\usepackage{nicefrac}   
\usepackage{float}

\newcommand{\sgrA}{Sgr A$^{\ast}$ }
\newcommand{\psr}{\textbf{[3sr]} }
\newcommand{\cyl}{\textbf{[uni]} }
\newcommand{\kpc}{\mathrm{kpc} }
\newcommand{\HESS}{H.E.S.S.\ }
\newcommand{\chired}{\chi^2_\mathrm{red}}
\newcommand{\diff}{\mathrm{d}}

\newcommand{\change}[1]{#1}

\shorttitle{anisotropic cosmic-ray transport in the Galactic Center}
\shortauthors{D\"orner et al.}

\begin{document}

\title{Impact of anisotropic cosmic-ray transport on the gamma-ray signatures in the Galactic Center}


\author[0000-0001-6692-6293]{J.~D\"orner}
\affiliation{Theoretische Physik IV, Fakult\"at f\"ur Physik \& Astronomie, Ruhr-Universit\"at Bochum, 44780 Bochum, Germany}
\affiliation{Ruhr Astroparticle and Plasma Physics Center (RAPP Center), Ruhr-Universit\"at Bochum, 44780 Bochum, Germany}

\author[0000-0002-1748-7367]{J.~Becker Tjus}
\affiliation{Theoretische Physik IV, Fakult\"at f\"ur Physik \& Astronomie, Ruhr-Universit\"at Bochum, 44780 Bochum, Germany}
\affiliation{Ruhr Astroparticle and Plasma Physics Center (RAPP Center), Ruhr-Universit\"at Bochum, 44780 Bochum, Germany}
\affiliation{Department of Space, Earth and Environment, Chalmers University of Technology, 412 96 Gothenburg, Sweden}

\author[0000-0002-5032-5896]{P.\ S.~Blomenkamp}
\affiliation{Ruhr Astroparticle and Plasma Physics Center (RAPP Center), Ruhr-Universit\"at Bochum, 44780 Bochum, Germany}
\affiliation{Astronomisches Institut, Fakult\"at f\"ur Physik \& Astronomie, Ruhr-Universit\"at Bochum, 44780 Bochum, Germany}

\author[0000-0002-9151-5127]{H.~Fichtner}
\affiliation{Theoretische Physik IV, Fakult\"at f\"ur Physik \& Astronomie, Ruhr-Universit\"at Bochum, 44780 Bochum, Germany}
\affiliation{Ruhr Astroparticle and Plasma Physics Center (RAPP Center), Ruhr-Universit\"at Bochum, 44780 Bochum, Germany}

\author[0000-0002-5605-2219]{A.~Franckowiak}
\affiliation{Ruhr Astroparticle and Plasma Physics Center (RAPP Center), Ruhr-Universit\"at Bochum, 44780 Bochum, Germany}
\affiliation{Astronomisches Institut, Fakult\"at f\"ur Physik \& Astronomie, Ruhr-Universit\"at Bochum, 44780 Bochum, Germany}

\author{E.\ M.~Zaninger}
\affiliation{Theoretische Physik IV, Fakult\"at f\"ur Physik \& Astronomie, Ruhr-Universit\"at Bochum, 44780 Bochum, Germany}
\affiliation{Ruhr Astroparticle and Plasma Physics Center (RAPP Center), Ruhr-Universit\"at Bochum, 44780 Bochum, Germany}


\begin{abstract}
The very high energy (VHE) emission of the Central Molecular Zone (CMZ) is rarely modelled in 3D. Most approaches describe the morphology in 1D or simplify the diffusion to the isotropic case.
In this work we show the impact of a realistic 3D magnetic field configuration and gas distribution on the VHE gamma-ray distribution of the CMZ.
We solve the 3D cosmic-ray transport equation with an anisotropic diffusion tensor using the approach of stochastic differential equations as implemented in the CRPropa framework. We test two different source distributions for five different anisotropies of the diffusion tensor, 
covering the range of effectively fieldline-parallel diffusion to isotropic diffusion.
Within the tested magnetic field configuration the anisotropy of the diffusion tensor is close to the isotropic case and three point sources within the CMZ are favoured.
Future missions like the upcoming CTA will reveal more small-scale structures which are not jet included in the model. Therefore, a more detailed 3D gas distribution and magnetic field structure will be needed.
\end{abstract}



\section{Introduction}
\input{01_introduction}

\section{Galactic Center environment} \label{sec:env}
\input{02_observations}

\section{Simulation setup} \label{sec:transport}
\input{03_transport} 

\section{Results} \label{sec:result}
\input{04a_synthetic_count_map}

\input{04b_profile}
\input{04c_spectra}

\input{04d_source_luminosity}

\section{Summary and Conclusion} \label{sec:summary}
\input{05_discussion}

\begin{acknowledgments}
    We acknowledge the support from the DFG via the Collaborative Research Center SFB1491  \textit{Cosmic Interacting Matters - From Source to Signal}. \\ 
    The authors gratefully acknowledge the computing time provided on
    the Linux HPC cluster at TU Dortmund University (LiDO3),
    partially funded in the course of the Large-Scale Equipment
    Initiative by the Deutsche Forschungsgemeinschaft (DFG, German
    Research Foundation) as project 271512359.
\end{acknowledgments}

\software{
 CRPropa \citep{CRPropa32}, dask \citep{dask}, ipython \cite{ipython}, matplotlib \citep{matplotlib}, numpy \citep{numpy}, pandas \citep{pandas} and scipy \citep{2020SciPy-NMeth}}



\appendix
\include{appendix}

\bibliography{references}{}
\bibliographystyle{aasjournal}

\end{document}

%% file: 01_introduction.tex
The Galactic Center (GC) is one of the most extreme and close-by astrophysical environments and of particular interest for studies of non-thermal processes.
The GC has been studied in all wavelengths from radio \citep{Heywood2022} to high- \citep{Ajello2016,DiMauro:2021raz} and very high energy (VHE) $\gamma$-rays \citep{HESS16, HESS18, Magic20, Veritas21}. 
The observed outflows at gamma-ray \citep{Ackermann14}, X-ray \citep{Sofue00}, microwave \citep{Finkbeiner04, Planck13} and radio wavelengths \citep{Pedlar89} as well as the small-scale structures like the non-thermal filaments and the molecular clouds \citep[see][for a review]{Henshaw23}
are in need of proper modelling.

The observation of VHE gamma-rays, first reported by the \textit{High Energy Stereoscopic System} \citep[H.E.S.S.,][]{HESS16}, hints towards the acceleration of cosmic-ray (CR) protons up to PeV energies. The GC is one of a few so-called \textit{PeVatrons} known in our Milky Way. The diffuse VHE gamma-ray emission in the GC has been spatially correlated with the dense gas of the central molecular zone (CMZ) and seems to be consistent with the injection of CRs by a steady state source located at the GC \citep{HESS18}. 
\change{In these papers}
the projected distribution of the H$_2$ column density inferred by the observation of the CS(1-0) line multiplied by a parameterised source profile of a Gaussian or a $1/r$-CR density profile \change{was used}.
This approach of modelling the CR transport neglects the existence of small-scale features in the 3D gas distribution as well as in the magnetic field. 

The first attempt to model the gamma-ray emission of the CMZ using a three-dimensional gas distribution 
was carried out by \cite{Scherer2022}. 
These authors probe whether the gas has an inner cavity or not. In lack of a three-dimensional magnetic field the authors assume isotropic diffusion for a Kraichnan spectrum of magnetic turbulence. Recently, these authors tested more realistic source distributions and a two-zone diffusion model \citep{Scherer2023}.

In this paper, we use the results of \cite{guenduez_bfield2020} in order to perform diffusive 3D CR propagation in the CMZ. We include the three-dimensional gas distribution for interactions and distinguish between parallel and perpendicular diffusion. Further, we test different source injection models, related to point sources inside the CMZ and a sea of Galactic CRs. 

The paper is organised as follows: in section \ref{sec:env} the Galactic Center environment and its observations are summarized. In section \ref{sec:transport} the transport model and simulation setup is presented and in section \ref{sec:result} the simulation results are compared to the observed data. Finally, in section \ref{sec:summary} a 
concluding discussion and an outlook is given.

%% file: 02_observations.tex
\subsection{VHE $\gamma$-ray observation}
The Galactic Center has been studied in VHE $\gamma$-rays ($>$ 100 GeV). 
The first detection of $\sim 100$~TeV $\gamma$-ray emission by \cite{HESS16} provided the first evidence for the existence of a \textit{PeVatron} in the Galactic Center region.
In \cite{HESS18}, \HESS quantified the spatial distribution of the diffuse $\gamma$-rays and the corresponding spectrum. Even MAGIC \citep{Magic20} and VERITAS \citep{Veritas21} have observed the CMZ. 
High energy $\gamma$-ray emission from the Galactic Center has been
detected at GeV energies by FermiLAT. As the GC region shows deviations from the typical expectation of cosmic-ray transport \citep{Ackermann2017}, Dark Matter has been proposed 
to be a possible contributor at GeV energies \citep{Goodenough2009, Daylan2016}. 

At $>100$~TeV $\gamma$-ray energies, LHASSO has reported on the detection of photons from the Galactic Plane \citep{Cao2023}. 
The IceCube collaboration also reported on the first observation of the Galactic Plane in high-energy neutrinos, which represents unambiguous proof for the signatures of hadronic CRs \citep{IceCube2023}. At this point, the exact contribution to the Galactic emission from the Galactic Center region cannot be quantified from the observational perspective. Theoretical studies like this paper can help to understand if a significant fraction of the total neutrino flux comes from the diffuse emission of the Galactic center, and we quantify the number of neutrinos that can be expected to come from the central, diffuse part in this study.

\subsection{Gas distribution}
The three-dimensional gas distribution of the CMZ is not well known, and the models are quite uncertain. 
We use the HI and H$_2$ CMZ components of the model by \cite{Ferriere07}. \change{It is parameterised as}
\begin{eqnarray}
        n(\vec{r}) = &n_0 \, \exp\left\{ - \left(\frac{\sqrt{X^2 + (2.5 Y)^2} - X_c}{L_c}\right)^4 \right\} \nonumber \\  
        &\times \, \exp\left\{ - \left(\frac{z}{H_c} \right)^2\right\} \quad,
\end{eqnarray} \\
\change{where $X$ and $Y$ are the local coordinates along major or minor axis.
} 

\begin{figure}
    \centering
    \includegraphics[width=\columnwidth]{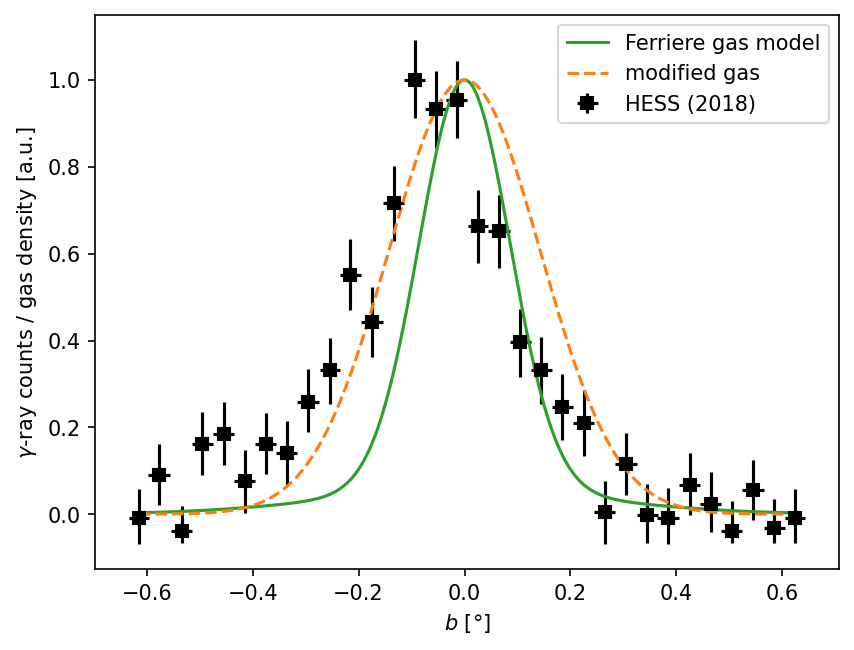}
    \caption{Comparison between the latitudinal profile ($b$) of the observed diffuse $\gamma$-ray flux (black squares) by \HESS \citep{HESS18} and model of the gas distribution (solid green line) by \cite{Ferriere07}. 
    The orange dashed line corresponds to the gas distribution using a larger parameter for the scale height of $H_c = 30\, \mathrm{pc}$.}
    \label{fig:gas}
\end{figure}

In contrast to the observed latitudinal profile of the diffuse $\gamma$-ray emission, the gas model shows a significantly 
shorter 
scale height of the disc (see Fig.~\ref{fig:gas}).
The maximal width of the latitudinal gamma-ray profile is, independently of the transport mechanism, determined by the gas distribution and the distribution suggested by \cite{Ferriere07} can not explain the observations.
Therefore, we adjust the scale height of the gas distribution to $H_c = 30 \, \mathrm{pc}$, which is near the upper limit of the observational uncertainties. \change{With this changes a reproduction of the latitudinal profile is possible (see section \ref{ssec:profiles} and Fig.~\ref{fig:profile}).}

\subsection{Magnetic field configuration}
To determine the local directions of parallel and perpendicular CR transport, the knowledge of the three-dimensional magnetic field configuration is crucial.
Here, we use the model proposed by \cite{guenduez_bfield2020}, which is a superposition of a large-scale intercloud (IC) component and more localised contributions. These small-scale components include the eight observed non-thermal filaments (NTF), 12 molecular clouds (MC), and a contribution from \sgrA. 
The IC and NTF components are predominantly poloidal while the molecular clouds are toroidal. In the MCs the ratio $\eta = B_r / B_\phi$ between the radial and azimuthal field is fixed to $\eta = 0.5$ as suggested in \cite{guenduez_bfield2020}.
The total magnetic field can be written as 
\begin{equation}
    \vec{B}_\mathrm{tot} = \vec{B}_\mathrm{IC} + \sum_{i=1}^{8} \vec{B}_{\mathrm{NTF}, i} + \sum_{i = 1}^{12} \vec{B}_{\mathrm{MC}, i}  + \vec{B}_\mathrm{SgrA^*} \quad .
\end{equation}

\begin{figure*}
    \centering
    \includegraphics[width=\textwidth]{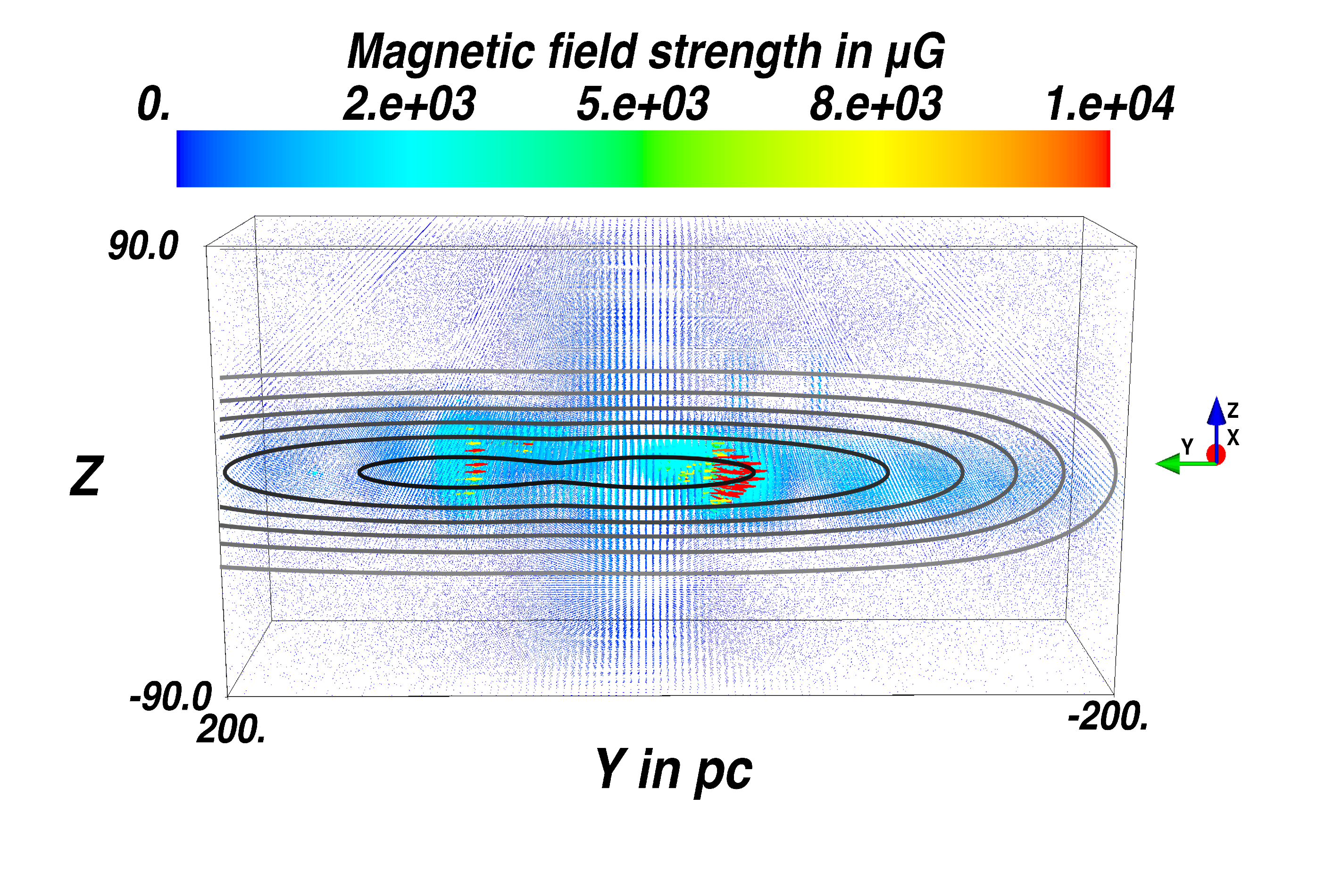}
    \caption{Superposition of the 3d magnetic field configuration from \cite{guenduez_bfield2020} [colored arrows] and the contours of the column density of the adapted gas distribution from \cite{Ferriere07} [grey lines].}
    \label{fig:field}
\end{figure*}

In Fig.~\ref{fig:field} a superposition of the magnetic field structure and the projected column density is given. \change{In general, the transport of CRs is determined by the inter-cloud component of the magnetic field. For the confinement of CRs and the local enhancement of gamma-ray emission the magnetic cloud sgr B2, the field around \sgrA and the NTF radio arc are the most important one. These structures are indicated in figure \ref{fig:sources} in cyan.}

%% file: 03_transport.tex
The transport of Galactic CRs can be described by the Parker equation (see, e.g., \citealt{Tjus2020})
\begin{equation}
    \frac{\partial n}{\partial t} = \nabla \hat{\kappa} \nabla n - \frac{\partial}{\partial E} \left[ \frac{\mathrm{d}E}{\mathrm{d}t} n\right] + S(\vec{r}, E, t)\quad , \label{eq:transport}
\end{equation}
where $n = n(\vec{r}, E, t)$ denotes the differential number density of CRs per unit volume, energy, and time. The diffusion tensor $\hat{\kappa}$ can be diagonalized in the frame of the local magnetic field line. Assuming the magnetic field is pointing in the $z$-direction $\vec{B} = B\vec{e}_z$ the diffusion tensor reads as $\hat{\kappa} = \mathrm{diag}(\kappa_\perp, \kappa_\perp, \kappa_\parallel)$. The details of the assumed diffusion tensor are given 
in sec.~\ref{ssec:diffusion}. The last term $S(\vec{r}, E, t)$ describes the sources and sinks of CRs, which are described in section \ref{ssec:sources}. 

The term $\mathrm{d}E/\mathrm{d}t$ quantifies the energy loss of CRs due to the interaction with the interstellar medium (ISM). In this process charged and neutral pions are produced, where the $\pi^0$ decay into two photons.  We use the hadronic interaction module presented in \cite{Hoerbe}, which is based on the parametrisation of the differential cross section in \cite{Kelner06}. 

We solve the transport equation using the method of stochastic differential equations (SDEs) as implemented in the public transport code CRPropa3.2 \citep{CRPropa3, Merten17, CRPropa32}. We calculate the 
steady-state solution following the approach in \cite{Merten17}. Details about the setup are given in section \ref{ssec:crpropa}.

\subsection{Diffusion tensor} \label{ssec:diffusion}
In general, the diffusive transport of CRs is anisotropic w.r.t.\ the local magnetic field line. This fact, originally discussed for the transport of CRs in the heliospheric magnetic field \citep{Jokipii-1966} and, subsequently, refined \citep[e.g.,][]{Effenberger-etal-2012, Shalchi-2021} as well as quantified \citep[e.g.,][]{Reichherzer2021b}, has in recent years also been acknowledged for their Galactic transport \citep[e.g.,][and references therein]{Effenberger-etal-2012b, Cerri-etal-2017}. This anisotropy is described by the diffusion tensor $\hat{\kappa}$ in the transport equation (eq.~\ref{eq:transport}). 

To quantify the anisotropy the ratio $\epsilon = \kappa_\perp / \kappa_\parallel$ between the diffusion coefficient perpendicular to the magnetic field line ($\kappa_\perp$) and along it ($\kappa_\parallel$) is used. In this work we consider five different values ($\epsilon = 10^{-3}, 10^{-2}, 0.1, 0.3, 1$) reaching from nearly purely parallel transport to isotropic diffusion. The value of this anisotropy should depend on the local turbulence and can vary spatially \citep{Reichherzer2020, Reichherzer2021, Reichherzer2021b}, but is not known for the GC. Therefore, we test different fixed values to show the impact of this anisotropy parameter. 

The energy-scaling for the diffusion coefficients is taken from quasi-linear theory and we normalise the parallel coefficient to match the observed value at Earth. With this, the parallel diffusion coefficient reads as 
\begin{equation}
    \kappa_\parallel(E) = 6.1 \cdot 10^{24} \, \frac{\mathrm{m^2}}{\mathrm{s}} \, \cdot \left(\frac{E}{4 \, \mathrm{GeV}}\right)^\frac{1}{3} \quad, 
\end{equation}
using the particle energy $E$. 

\subsection{Sources} \label{ssec:sources}
\begin{figure}
    \centering
    \includegraphics[width=1.0\columnwidth]{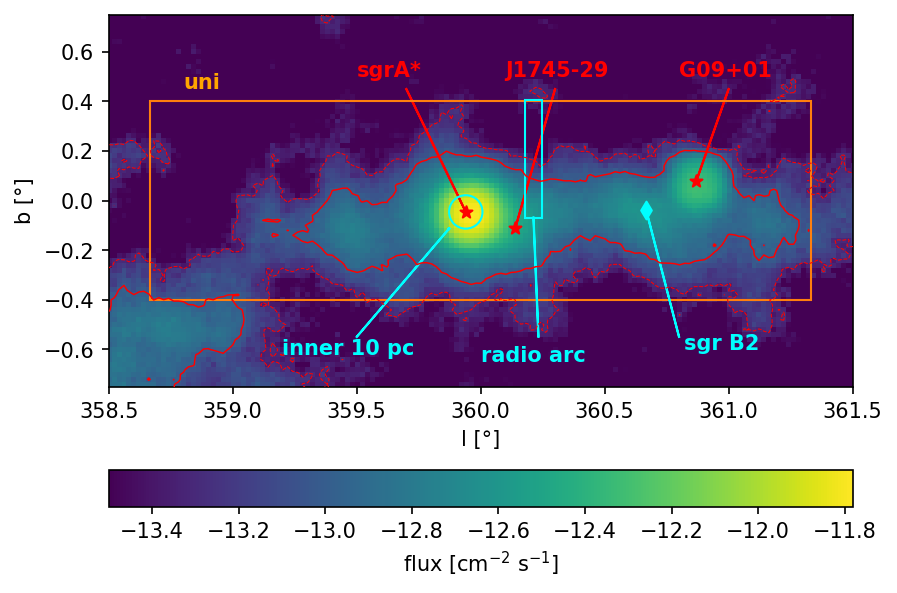}
    \caption{The VHE $\gamma$-ray flux from the CMZ observed by \HESS \cite{HESS18}. The red lines is the contour for $\Phi = 10^{-13} \mathrm{cm^{-2} \, s^{-1}}$ (red solid) and $\Phi = 5\cdot 10^{-14} \mathrm{cm^{-2}\, s^{-1}}$ (red dashed). The three red stars indicate the position of CR sources. The orange box indicates the simulation volume at $x = 0$. In cyan the three most important magnetic field components (the inner 10 pc structure, the radio arc and the molecular cloud sgr B2) are marked.}
    \label{fig:sources}
\end{figure}

As the origin of CRs is not clear we test two different scenarios for the spatial source distribution:
\begin{itemize}
    \item[\textbf{[3sr]}] The first source scenario considers the three observed $\gamma$-ray point sources as observed by \HESS These are (1) the central source  \sgrA (also called HESS J1745-290), (2) the supernova remnant G0.9+01 and (3) the pulsar HESS J1746-28.
    In this scenario, the contribution of the individual sources to the total CR luminosity is based on the $\gamma$-ray observation in \cite{HESS18}. This corresponds to a fraction of $f_\mathrm{sgrA} = 0.72$, $f_\mathrm{G0.9} = 0.22$ and $f_\mathrm{J1746} = 0.06$.
    \change{A comparison between the source positions in this work and those used in \cite{Scherer2022} is given in appendix \ref{app:sources}.}
    
    \item[\textbf{[uni]}] In the second source scenario, the full simulation volume is filled by a homogeneous CR source. This distribution could correspond to a CR population which is accelerated outside the GC region and diffused in a long time ago. 
\end{itemize}
An overview of the source position is indicated in fig.~\ref{fig:sources} by the red stars for the \psr scenario and the orange rectangle for the \cyl source scenario. 

In this work, we restrict our model to only contain protons. These are injected with a flat power-law spectrum $\left. \nicefrac{\mathrm{d}N}{\mathrm{d}E}\right|_\mathrm{sim} \sim E^{-1}$ to ensure equal statistics in each logarithmic energy bin. In the post-processing of the simulation data, the source spectrum is re-weighted to a power law 
$\left. \nicefrac{\mathrm{d}N}{\mathrm{d}E}\right|_s \sim E^{-\alpha_s}$ 
by assigning a weight 
\begin{equation}
    w_i = \frac{E_i^{\alpha_s - 1}}{\sum_i E_i^{\alpha_s - 1}}
\end{equation}
to each pseudo-particle,
which is called \textit{candidate} in CRPropa,
as presented in \cite{Merten17}. 

\subsection{CRPropa configuration} \label{ssec:crpropa}
The simulation volume is a paraxial box of the size $200 \times 400 \times 120$ pc$^3$ centered on \sgrA. For each source configuration and anisotropy parameter, a set of 50 simulations with $N_\mathrm{sim} = 10^5$ primary CRs is 
performed. 
This splitting is necessary to keep the simulation time per run as well as the amount of data managable \footnote{This corresponds to $\sim 600$ CPU-h computation time and $\sim 2$ GB data output per run.}.

The details of the used modules for the simulations are summarised in table \ref{tab:CRP_modules}. The output contains all created $\gamma$-rays directly after their production. No propagation and corresponding absorption of $\gamma$-rays is taken into account. To realise this, we use the \texttt{DetectAll} observer and set a veto for nucleons.

\renewcommand{\arraystretch}{1.5}
\begin{table*}[ht]
    \centering
    \begin{tabular}{|c|c|c|}
        \hline
         module & parameter & value  \\ \hline
         \multicolumn{3}{|c|}{magnetic field \& propagation} \\  \hline
         \texttt{CMZField} & sub-components & \texttt{True} \\
         \texttt{DiffusionSDE} & precision & $P = 10^{-3}$ \\
         & minstep & $s_\mathrm{min} = 10^{-3} \, \mathrm{pc}$ \\
         & maxstep & $s_\mathrm{max} = 10 \, \mathrm{pc}$ \\ 
         & anisotropy & $\epsilon \in \left\{ 10^{-3}, 10^{-2}, 0.1, 0.3, 1 \right\}$ \\ 
        \hline \multicolumn{3}{|c|}{observer \& output} \\ \hline 
         \texttt{HDF5Output} & enabled columns & TrajectoryLength \\
         & & position (source and current) \\
         & & energy (source and current) \\
         & & serial number  \\
         \texttt{Observer} & Particle veto & nucleus, electron, neutrino \\
         & Observer feature & \texttt{ObserverDetectAll} \\
        \hline \multicolumn{3}{|c|}{boundary \& break condition} \\ \hline
        \texttt{MaximumTrajectoryLength} & maximal time & $T_\mathrm{max} = 500 \, \mathrm{kpc}/c$ \\ 
        \texttt{MinimumEnergy} & minimal energy & $E_\mathrm{min} = 1 \, \mathrm{TeV}$ \\
        \texttt{ParaxialBox} & origin & $ \vec{o} = (-100, -200, -60) \,\mathrm{pc}$\\
        & size & $\vec{s} = (200, 400, 120) \, \mathrm{pc}$ \\
        \texttt{ObserverSurface} & surface & paraxial box as defined before. \\
        \hline \multicolumn{3}{|c|}{source} \\ \hline 
        \texttt{SourceParticleType} & particle id & proton ($1000010010$) \\ 
        \texttt{SourceIsotropicEmission} & & \\
         \texttt{SourceMultiplePositions} & positions & \sgrA : $\vec{r} = (0, 8.9, -6.8) \, \mathrm{pc}$ \\
         & & J1746: $\vec{r} = (0, -20.77, -16.32)\, \mathrm{pc}$ \\ 
         & & G0.9+01: $\vec{r} = (0, -129.08, 11.87) \, \mathrm{pc}$ \\
         or \texttt{SourceUniformBox} & origin / size & $\vec{o}$ and $\vec{s}$ as above
     \\ \hline    \end{tabular}
    \caption{CRPropa modules used for the simulation and their input parameters. 
    Module parameters not mentioned are kept at their default values.}
    \label{tab:CRP_modules}
\end{table*}

The \texttt{DiffusionSDE} module \citep[see][for details]{Merten17} is used to calculate the solution of the transport equation. We use an adaptive step size with a 
precision of $P = 10^{-3}$. The diffusion tensor is described in sec.~\ref{ssec:diffusion}. To speed up the simulation in the case of isotropic diffusion, we use a uniform magnetic field in the $z$-direction. In this case, the transport does not depend on the magnetic field configuration, but the adaptive step size method would lower the steps to resolve the curvature of the magnetic field. 

We limit the simulation to primary particles with a minimal energy of $E_\mathrm{min} = 1 \, \mathrm{TeV}$ and a maximum simulation time 
\footnote{Please note that the maximal simulation time is chosen to be much longer than the typical time a pseudo-particle spends in the simulation volume. In appendix \ref{app:sec:particleLeakage} it can be seen, that the particles leave the simulation volume earlier.}
of $T_\mathrm{max} = 500\, \mathrm{kpc}/{c}$ . Moreover, all particles reaching the boundary of the simulation volume are lost.

\subsection{Post processing} \label{ssec:postproc}
After a simulation, all produced $\gamma$-rays are binned and reweighted according to the primary energy. This is done for different power-law indices $\alpha_s$ of the source emission. We test $1 \leq \alpha_s \leq 3$ with steps of $\Delta \alpha_s = 0.1$. 
The data are binned in longitude, latitude, and energy. In the first step, the binning is done in a much finer resolution than current generation imaging Air Cherenkov telescopes (IACTs) can resolve. We use an angular binning of $\Delta l = 0.016^\circ$ and $\Delta b = 0.01^\circ$ to ensure enough statistics in each bin. The resolution effects of the observation are later taken into account by smearing the results. This allows us to compare the data for future telescopes like the upcoming CTA, which will have a two to three times better resolution \citep{CTA_Science}. The energy binning is done in the same ranges as in the \HESS analysis \citep{HESS18}. 

%% file: 04a_synthetic_count_map.tex
\subsection{Synthetic count maps}
\begin{figure*}[htp]
    \centering
    \includegraphics[width=\textwidth,height=\textheight,keepaspectratio]{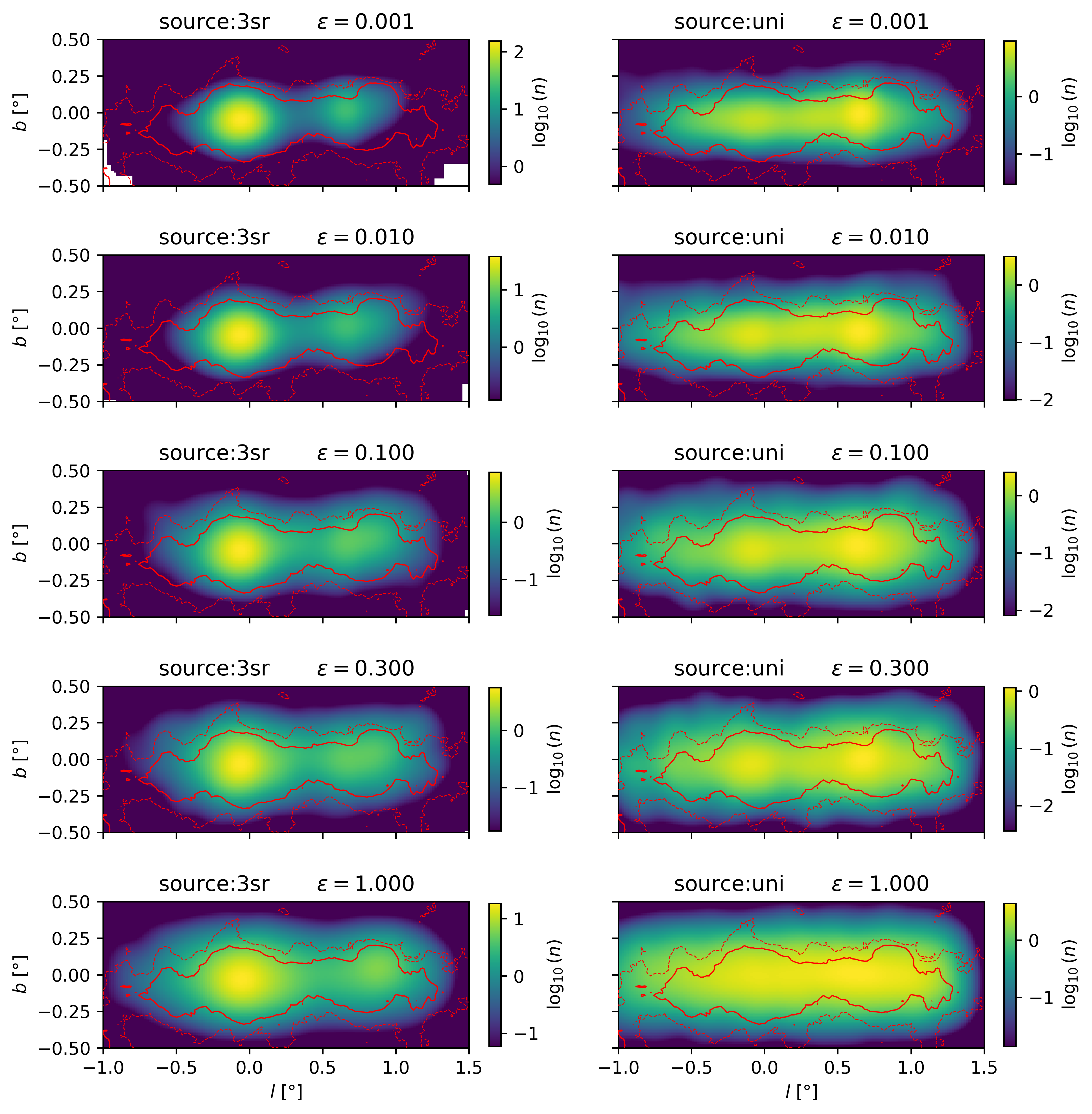}
    \caption{Synthetic $\gamma$-ray maps for a source injection with $\alpha_s = 2.0$. The first column shows the \psr source distribution and the second column the \cyl injection scenario. The row denotes the anisotropy parameter $\epsilon \in \{0.001, 0.01, 0.1, 0.3, 1\}$. The thin red lines show the contours of the observed flux by HESS \citep{HESS18_GPS} for $\Phi = 5\cdot 10^{-14}\, \mathrm{cm^{-2}\,s^{-1}}$ (dashed) and  $\Phi = 10^{-13} \, \mathrm{cm^{-2}\,s^{-1}}$ (solid).}
    \label{fig:2d_countmaps}
\end{figure*}

Using the produced photons in our simulation, we create synthetic $\gamma$-ray maps by calculating weighted histograms of the longitudinal and latitudinal positions. 
The weighting takes the injection spectrum $\left. \nicefrac{\mathrm{d}N}{\mathrm{d}E} \right|_s \sim E^{-\alpha_s}$ into account. 
The resulting synthetic count maps for a source index $\alpha_s = 2.0$ are shown in Fig.~\ref{fig:2d_countmaps}. In general the maps do not change significantly for different source spectra. To compare our simulation with the observations by \HESS we apply a Gaussian smearing of $\sigma = 0.077^\circ$, which is the 68\% containment radius of the point spread function \citep{HESS18}. 
Appendix \ref{app:sec:countmap} contains the same countmaps for the raw data without smearing.

In the case of strong parallel diffusion ($\epsilon = 0.001$) the CRs mainly follow the magnetic field lines. This leads to stronger confinement of CRs in the MCs around \sgrA and Sgr B2. Therefore, also the $\gamma$-ray production is centred near the sources/MCs.  For the \psr source scenario, the emission around \sgrA is stronger as the two sources close by emit more CRs. The emission in Sgr B2 shows the direction of the local field line where the CRs diffuse. 
By increasing perpendicular diffusion the point-like emission is smeared out. The production of $\gamma$-rays at higher latitudes becomes more likely and the distribution across the plane is spread out. In the extreme case of isotropic diffusion ($\epsilon = 1$) the point source G0.9 is barely visible and hidden by the large-scale diffuse emission. 

The $\gamma$-ray maps for the \cyl source injection show an extended disk for all anisotropies. This is expected due to the extended source distribution. Although the sources are distributed homogeneously, a concentration of produced photons around \sgrA and Sgr B2 is visible. This effect is mainly caused by the stronger magnetic fields in these regions. As these fields are much more twisted the confinement of CRs is more efficient, which also leads to a higher $\gamma$-ray production rate. Especially in the parallel diffusion-dominated case ($\epsilon = 10^{-3}$) the strongest emission is centred in Sgr B2. 
For the \cyl source injection the increasing isotropy leads to a wider spread of $\gamma$-rays in latitude, while no clear difference between the longitudinal profiles is visible. Only in the case of isotropic diffusion the peaks of \sgrA and Sgr B2 are not visible anymore, as this scenario does not depend on the magnetic field configuration.

%% file: 04b_profile.tex
\subsection{Count profiles} \label{ssec:profiles}
To quantify the difference between the distribution of photons we compare our calculated $\gamma$-ray maps with the latitudinal and longitudinal profile presented by H.E.S.S.\  \citep[see Fig.~4 of][]{HESS18}. Analogously to the H.E.S.S. data analysis all $\gamma$-rays in a latitudinal (longitudinal) window of $|b| \leq 0.3^\circ$ ($|l| \leq 0.5^\circ$) are collected. The profiles are calculated on a much finer binning ($\Delta l = 0.016^\circ$ and $\Delta b = 0.01^\circ$) and smeared with the H.E.S.S. resolution of $\sigma = 0.077^\circ$. 
The simulation data are normalised to match the maximal counts of the latitudinal profile for $b = -0.054^\circ$, which is the middle of its peak.

\begin{figure*}[hp]
    \centering
    \includegraphics[width=\textwidth,height=.95\textheight,keepaspectratio]{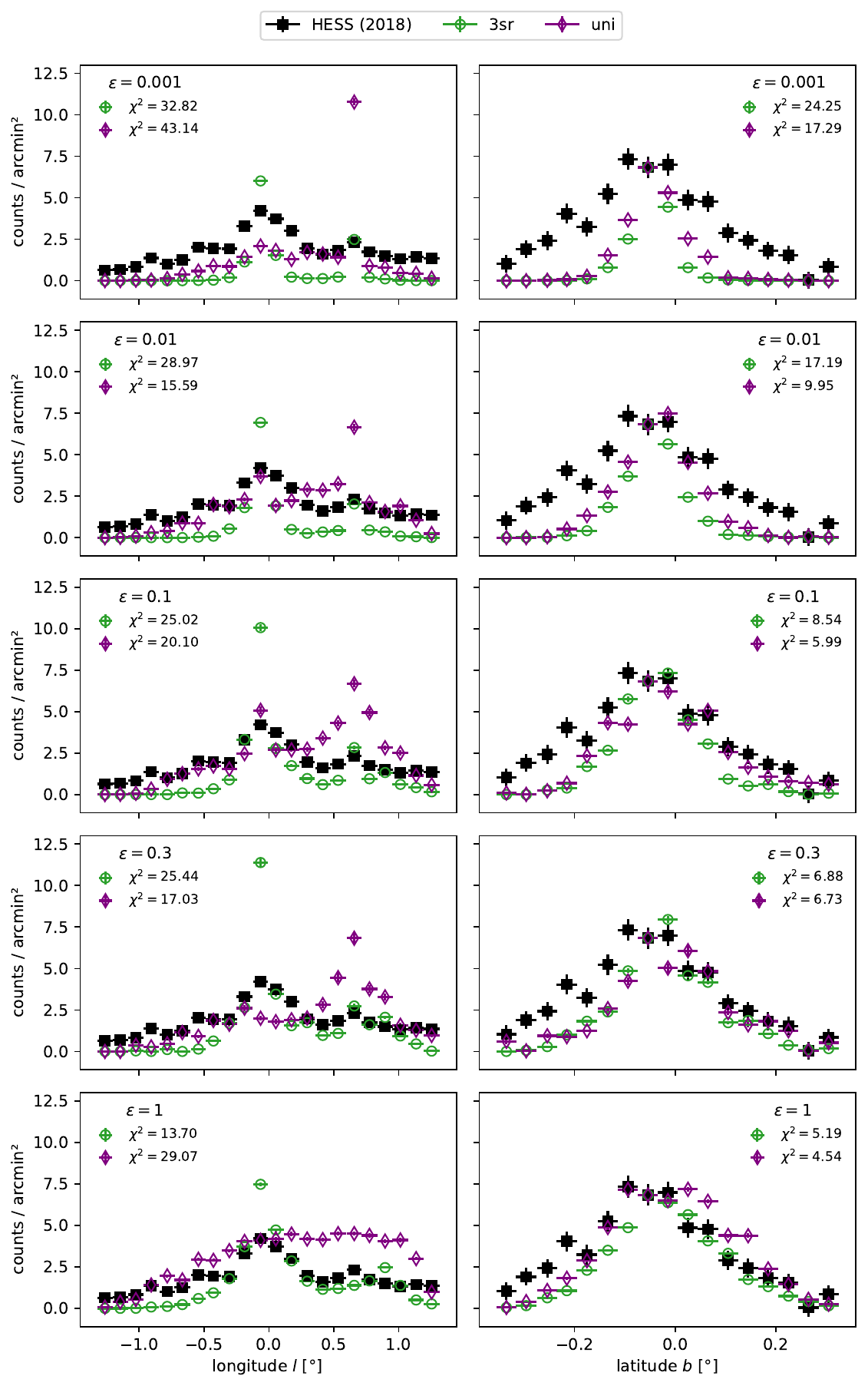}
    \caption{Profiles of the $\gamma$-ray distribution along the longitude (left column) and the latitude (right column). The row denotes the anisotropy parameter.}
    \label{fig:profile}
\end{figure*}

In Fig.~\ref{fig:profile} the profiles for a source injection with a power law of $\left. \nicefrac{\mathrm{d}N}{\mathrm{d}E} \right|_s \sim E^{-2}$ is shown. The difference for varying the power law slope $\alpha_s$ is shown in appendix \ref{app:sec:source_spectrum}. To estimate the agreement between the data and the simulation the reduced $\chi^2$ 
\begin{equation}
    \chi^2_\mathrm{red} = \frac{1}{n - 1} \sum_{i=1}^{n} \frac{\left( c_i^{(\mathrm{obs})}  - c_i^{(\mathrm{sim})}\right)^2}{\sigma_i^2} \label{eq:chi2}
\end{equation}
is calculated. Here $c_i$ is the observed or simulated number of counts, $\sigma_i$ is the observational uncertainty and $n$ is the total number of data points. 

The latitudinal profiles show for both source scenarios a much too thin disk for high anisotropies of the diffusion tensor ($\epsilon = 0.001$ or $\epsilon = 0.01$).
For the \cyl source model, the latitudinal profile is matched best using $\epsilon = 0.1$, while the \psr model prefers $\epsilon = 0.3$. But in both source scenarios, the anisotropic diffusion is favoured over the isotropic case ($\epsilon = 1$). 

In contrast to the latitudinal profile, where both source scenarios show the same shape, clear differences are visible in the longitudinal profiles (left column of Fig.~\ref{fig:profile}). For the smallest anisotropy, the differences are most dominant. The \psr model shows a 
considerable peaking around the positions of the sources and nearly no $\gamma$-ray production further away. This is expected due to the strong confinement of CRs in the local environment of the sources. The \cyl model shows a nearly smooth distribution over the full range. Only at the position $l = 0.657^\circ$ a maximum is visible. This can be explained by the strong magnetic field of the MC Sgr B2. 
With stronger perpendicular diffusion the peak of the \psr model is broader. For the \cyl source model, the trend is the opposite way. In the case of stronger perpendicular diffusion (up to $\epsilon = 0.1$) the longitudinal profile gets higher around \sgrA, while the peak at Sgr B2 gets spread out. 
This shift of the peaks in the $\gamma$-ray distribution might result from the small-scale structure of the magnetic field. Especially the position of the MCs and NTFs along the line of sight has a strong impact on the confinement of CRs. 
Overall the strongest impact on the gamma-ray distribution by small-scale magnetic field is given by the MC sgr B2 and the field around \sgrA. Those components are highlighted in Fig.~\ref{fig:sources}.

\begin{figure}[ht]
    \centering
    \includegraphics[width=\columnwidth]{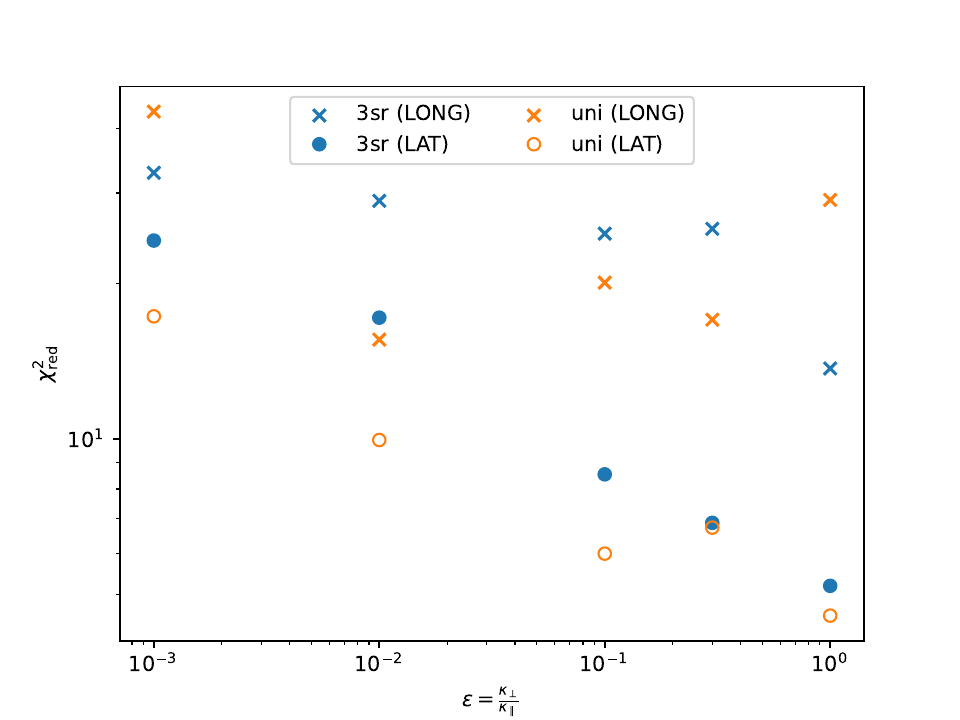}
    \caption{Reduced $\chi^2$ values to quantify the difference between the observed count profiles and the simulation (compare Fig.~\ref{fig:profile}).}
    \label{fig:chi2_count}
\end{figure}

A comparison between the $\chi^2_\mathrm{red}$ values depending on the anisotropy of the diffusion tensor is shown in Fig.~\ref{fig:chi2_count}. 
In the latitudinal comparison the $\chired$ is comparable for all anisotropies $\epsilon \gtrsim 0.1$ and both source distributions. This is expected due to the fixed normalisation of the count rate and only the width of the distribution can change with the anisotropy.

Comparing the longitudinal profiles offers a better discrimination between the source models and anisotropies, as the normalisation is followed from the latitudinal profile.
Taking only the longitude into account the best agreement to the data is reached for the \psr source distribution assuming isotropic diffusion ($\epsilon = 1$), although it still overshoots the height of the central peak. 
All \cyl models over-predict the peak at Sgr B2 and under-predict the peak at \sgrA and can be ruled out.

%% file: 04c_spectra.tex
\subsection{Spectra} \label{ssec:SED}
Additional to the angular distribution of the $\gamma$-ray flux, also the spectral energy distribution (SED) is measured by \HESS \citep{HESS16}, MAGIC \citep{Magic20} and VERITAS \citep{Veritas21}. 
For this analysis, we consider different slopes of the CR injection spectrum $\left. \nicefrac{\mathrm{d} N }{\mathrm{d} E}\right|_s \sim E^{-\alpha_s}$. Indices in the range $1 \leq \alpha_s \leq 3$ with steps of $\Delta \alpha_s = 0.1$ are tested. 
For all configurations (source distribution, anisotropy of the diffusion tensor and source injection index) we bin the simulation data according to the \HESS observation, as they provide the finest energy resolution. The simulated SED is fitted with a power law 
\begin{equation}
    \Phi = \frac{\mathrm{d}N}{\mathrm{d}E} =  \Phi_0 \, \left( \frac{E}{1 \, \mathrm{TeV}} \right)^{- \alpha} \quad, \label{eq:powerlaw}
\end{equation}
where $\Phi_0$ is the normalisation at 1 TeV and $\alpha$ is the spectral index and a power-law with exponential cut-off
\begin{equation}
    \Phi = \Phi_0 \, \left( \frac{E}{1 \, \mathrm{TeV}} \right)^{-\alpha} \, \exp\left\{ - \frac{E}{E_c} \right\} \label{eq:cutoff}
\end{equation}
with the cut-off energy $E_c$. The simulations allow in both cases for a free normalisation $\Phi_0$. Therefore, we choose the normalisation to minimise the $\chi^2$ difference between the fit and the observed data.

\begin{figure}[ht]
    \centering
    \includegraphics[width=\columnwidth]{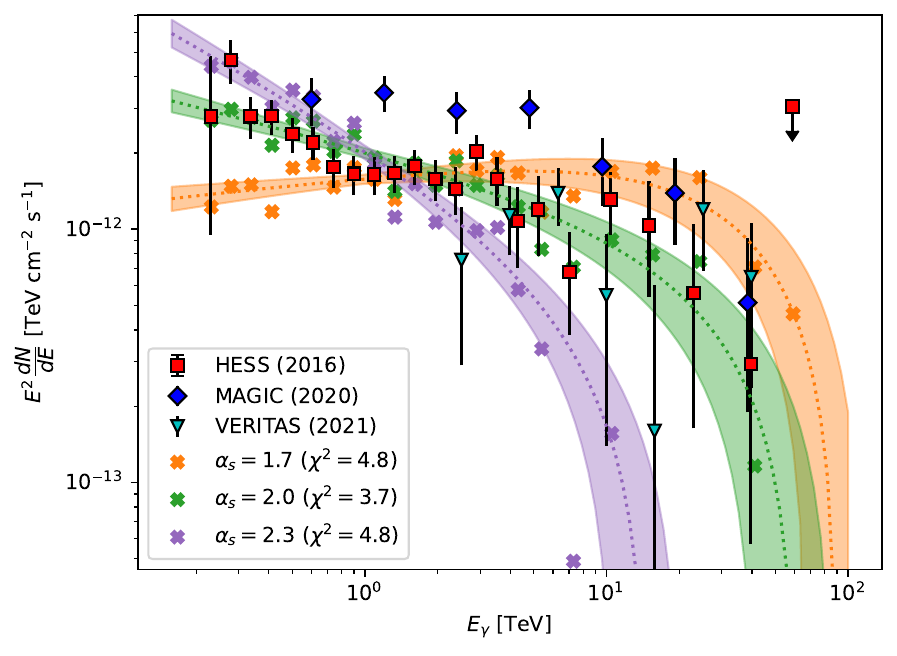}
    \caption{Example SED for the \psr source distribution and $\epsilon = 0.1$. Three different injection slopes $\alpha_s$ are shown.}
    \label{fig:SED_example}
\end{figure}

\renewcommand{\arraystretch}{1.2}
\begin{table*}[!ht]
    \begin{tabular}{c|ccc|ccc}
         source & $\chi^2_\mathrm{red}$ & $\epsilon$ & $\alpha_s$ & $\log_{10} \left(\Phi_0 [\mathrm{TeV\, cm^{-2}\, s^{-1}}]\right)$& $\alpha$ &$E_c$ [TeV] \\ \hline
         \psr & $3.23$ &$0.3$ & $1.9$ & $-11.705 \pm 0.022$ & $2.219 \pm 0.028$ & - \\
         & $3.65$ & $0.1$ & $2.0$ & $-11.692 \pm 0.014$ & $2.21 \pm 0.03$ & $79 \pm 21$ \\
         \cyl & $3.23$ &$0.001$ & $1.9$ & $-11.702 \pm 0.019$ & $2.233 \pm 0.024$ & - \\
          & $3.37$ &$0.1$& $2.0$ & $-11.695 \pm 0.008$& $2.250 \pm 0.016$ & $307 \pm 169$\\
    \end{tabular}
    \caption{Best fit parameter for the minimal $\chi^2_\mathrm{red}$ shown in Fig.~\ref{fig:chi2_SED}. For both spatial models the first row corresponds to the power-law fit (eq.~\ref{eq:powerlaw}) and the second row to the cut-off fit (eq.~\ref{eq:cutoff}).}
    \label{tab:best_fit}
\end{table*}

\begin{figure*}[ht]
    \centering
    \includegraphics[width=.8\textwidth]{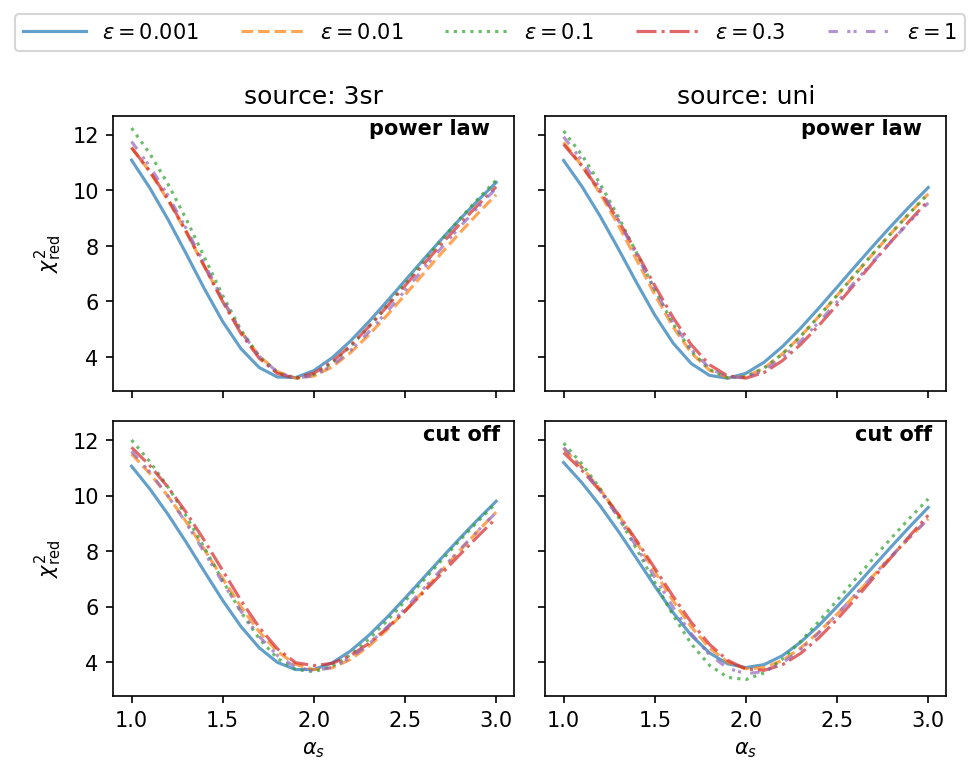}
    \caption{$\chi^2$ difference between the fitted SED and the observation depending on the source injection slope. The column indicates the spatial source distribution (left \psr, right \cyl) and the row indicates the fitting function (\textit{upper}: power law (eq.~\ref{eq:powerlaw}), \textit{lower}: power law with cut-off (eq.~\ref{eq:cutoff})). The line color and style denote different anisotropies of the diffusion tensor.}
    \label{fig:chi2_SED}
\end{figure*}

In Fig.~\ref{fig:SED_example} the SED is shown for the injection slope of $\alpha_s = 1.7$ (orange), $\alpha_s = 2.0$ (green) and $\alpha_s = 2.3$ (purple) assuming the \psr spatial distribution and the anisotropy parameter of $\epsilon = 0.1$.  
Also the observation by \HESS \citep{HESS16} (red squares), MAGIC \citep{Magic20} (blue diamond) and VERITAS \citep{Veritas21} (cyan triangle) are shown.

The results for the minimised $\chi^2$ based on the different injection slopes is shown in Fig.~\ref{fig:chi2_SED}.
All cases lead to the best fit for an injection slope $\alpha_s \approx 2$, while the pure power-law fit requires a slightly harder injection spectrum. The parameters for the best fits are summarised in table \ref{tab:best_fit}. No clear preference for the anisotropy parameter can be found for the SED fitting.

Due to the hadronic nature of the gamma-ray production, we also expect a neutrino component. Neglecting the absorption of gamma-rays the all flavour neutrino flux can be approximated as $\Phi_{\nu, tot}(E_\nu) = 6 \, \Phi_\gamma(E_\nu / 2)$ \citep{Tjus2020}. 
Using the neutrino flux $\Phi_\nu$ and the effective area $A_\mathrm{eff}$ of the neutrino detection from the Galactic Plane \citep{IceCube2023} we can calculate the expected number of neutrinos within the 10 years of IceCube data as 
\begin{equation}
    N_\nu = \Delta t_\mathrm{obs} \, \int \Phi_\nu(E) \, A_\mathrm{eff}(E) \, \mathrm{d}E \quad.
\end{equation}
In the most optimistic scenario, a powerlaw-like emission from the \psr source distribution, the expected number of neutrinos is $N_\nu = 0.016$. This makes the observation of the CMZ as a neutrino point source within the Galactic Plane very unlikely.

%% file: 04d_source_luminosity.tex
\subsection{\change{Source luminosity}} \label{ssec:luminosity}
\change{
To estimate the required energy budget of the CR sources within the CMZ we ran a set of smaller simulations with $N_\mathrm{sim} = 10^4$ for each anisotropy $\epsilon$. Additionally to the simulation setup described in section \ref{ssec:crpropa}, an output of all injected primary protons is added. 
}

\change{
The absolute normalisation of the simulated CR and $\gamma$-ray fluxes can be achieved by assigning weights to \textit{candidates} based on the luminosity at the source $L_\mathrm{src}$ and the distance to the observer $r_\mathrm{obs}$ \citep[see, e.g appendix A in][]{Eichmann2023}. 
In our case, considering only one source species and assuming all photons to be emitted at $r_\mathrm{obs} = 8.5\, \mathrm{kpc}$, the normalisation factor between the simulated flux $J$ (in units particles / TeV) and the physical flux $\Phi$ reduces to a constant factor and is independent of the observed particle species. 
}

\change{
To estimate the normalisation factor 
\begin{equation}
    f^\gamma = \frac{J_\gamma (1 \, \mathrm{TeV})}{\Phi_\gamma (1\, \mathrm(TeV)}
\end{equation} 
we perform the same fitting as described in section \ref{ssec:SED} on the smaller test setup assuming a source injection with $\alpha_s = 2.0$. Applying the same factor to the simulated proton spectrum at the source $J_p$ we can estimate total CR luminosity as  
\begin{equation}
L_p (E \geq 10 \, \mathrm{TeV}) \approx \int\limits_{10 \, \mathrm{TeV}}^{1 \, \mathrm{PeV}} \diff E_p \, E_p \, J_p(E_p) \, f^\gamma \, 4\pi r_\mathrm{obs}^2 \quad .
\end{equation}
}

\begin{figure}
    \centering
    \includegraphics[width=\columnwidth]{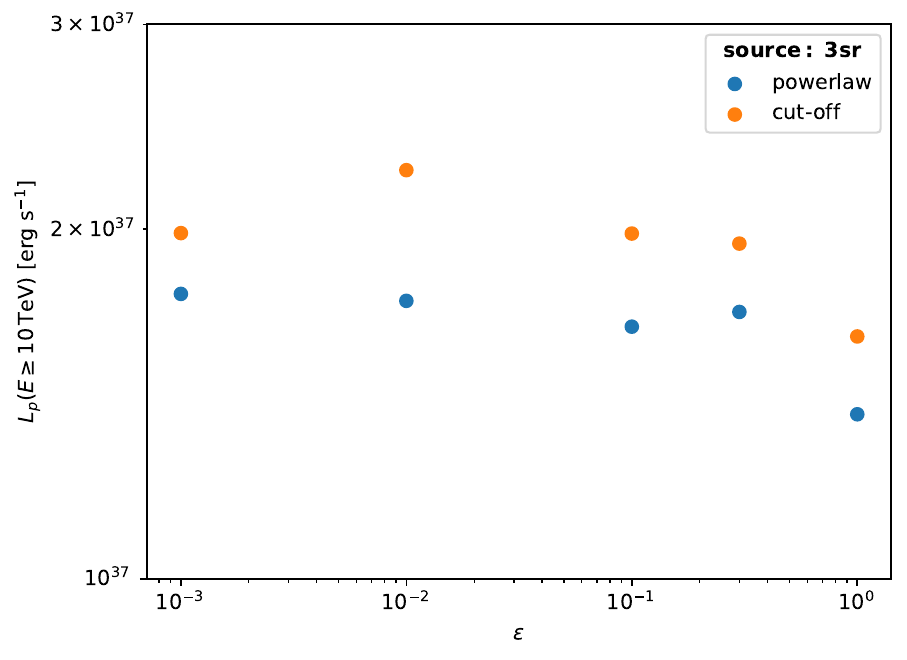}
    \caption{Estimated source luminosity in the \psr source scenario, assuming $\alpha_s = 2.0$}
    \label{fig:luminosity}
\end{figure}

\change{
The resulting source luminosity, depending on the anisotropy and the type of the performed fit is shown in Fig.~\ref{fig:luminosity}. The source luminosity depends only slightly on the assumed anisotropy of the diffusion tensor and is within the required acceleration rate of $10^{37} - 10^{38} \, \mathrm{erg}/\mathrm{s}$ as claimed by \HESS \citep{HESS16}.  
}

%% file: 05_discussion.tex
Building a realistic three-dimensional model of the cosmic-ray (CR) transport inside the Central Molecular Zone (CMZ) requires detailed knowledge of the astrophysical environment, i.e. the gas distribution,  the magnetic field configuration, and the source positions. 

In our work, we used the three-dimensional gas distribution from \cite{Ferriere07}. We adjusted the exponential scale height to $H_c = 30 \ \mathrm{pc}$ to match the observed thickness of the $\gamma$-ray emission, which is close to the upper limit of the observational uncertainties. All transport scenarios using the original thickness would lead to a thinner disk, as the maximal width of the gamma-ray distribution is given by the density distribution. Increasing the scale height to even higher values would also allow the more anisotropic diffusion scenarios to match the latitudinal profile, but the longitudinal behaviour would not be affected.

The anisotropy of the diffusion tensor $\epsilon = \kappa_\perp / \kappa_\parallel$, defined as the ratio between the diffusion perpendicular and parallel to the magnetic field line, is constrained by the observation of the longitudinal and latitudinal profiles and the SED of the $\gamma$-ray emission. The measurements by the \HESS telescopes \citep{HESS18} hint towards a nearly isotropic diffusion of CRs, while in the SED fitting no clear preference can be seen. \change{The required source luminosity is reasonable within the expected range by \cite{HESS16}.}

In this work, we tested two different source scenarios, three different point sources within the CMZ, and a global sea of old CRs from the Milky Way diffusing into the CMZ. 
The discrimination between the different source distributions and anisotropies is done best by comparing the longitudinal profiles. In this case, the best agreement with the data could be achieved by the point source scenario. Here, the smallest $\chired$ can be achieved. Only the position of the peak for positive longitudes is shifted a little bit outward. The distribution underpredicts the outermost part. This might hint towards a missing gas target in this range or a contribution from the CR sea.

In general, the large-scale observables could be reproduced with our 3D model of the CMZ but some of the small-scale features
like the small enhancement at $l \lesssim -0.5^\circ$ and the exact position of the peak around $l \sim 0.7^\circ$
are still missing. This is mainly due to the lack of substructures within the gas distribution, but also a more refined magnetic field model would be needed. 
In the outer part of the model the transition between the dense CMZ gas and the thinner Galactic disk needs to be modelled more carefully.

This will become important with the upcoming next-generation telescopes like CTA, which will provide a lower angular resolution and a better sensitivity.
It will allow us to distinguish between the contribution from different molecular clouds and the improved statistic will enable us to test spectral differences between different regions within the CMZ.

%% file: appendix.tex
\section{Comparing source positions}
\label{app:sources}
\change{
In this section the differences of the source positions used in this work to those used by \cite{Scherer2022} is shown. \cite{Scherer2022} focused on the emission from star clusters (Nuclear Star Cluster [NSC], Arches Cluster [AC] and Quintuplet Cluster [QC]). 
Additionally they introduce the source SgrA east as an impulsive source. The distances of NSC and sgrA east to our source SgrA* are much shorter than the resolution of H.E.S.S. Therefore, we do not expect a difference here. }

\change{The other clusters (AC and QC) are a possible source for cosmic-ray acceleration but are not observed in VHE gamma-ray. Based on this we use the identified gamma-ray source HESS J1745-29 instead. In the outer part Scherer et al
identified the SNR sgrD as contributing source. They discuss three G-objects as possible source (G1.1–0.1, G1.0–0.2 and G0.9+0.1). They arbitrarily choose the middle one G1.0-0.2 as source, but only G0.9+0.1 is observed in VHE
gamma-ray. Therefore, we use this source. }

\begin{figure}
    \centering
    \includegraphics[width=\columnwidth]{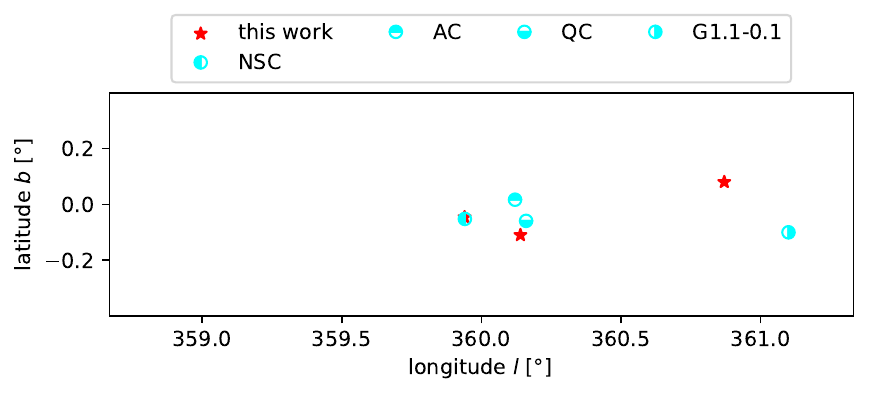}
    \caption{Comparison of the source positions in this work (red stars) to those used by \cite{Scherer2022} (cyan circles).}
    \label{app:fig:sources}
\end{figure}

\change{In figure \ref{app:fig:sources} the different source position from this work and \cite{Scherer2022} are shown.}

\section{Estimating the maximal simulation time} \label{app:sec:particleLeakage}
To estimate the necessary total simulation time we test the number of particles left in the simulation volume in steps of $\Delta t = 20 \, \frac{\kpc}{c}$. This is done for all simulation setups with $N_\mathrm{sim} = 10^4$ primaries. 

\begin{figure}[htb]
    \centering
    \includegraphics[width=.9\columnwidth]{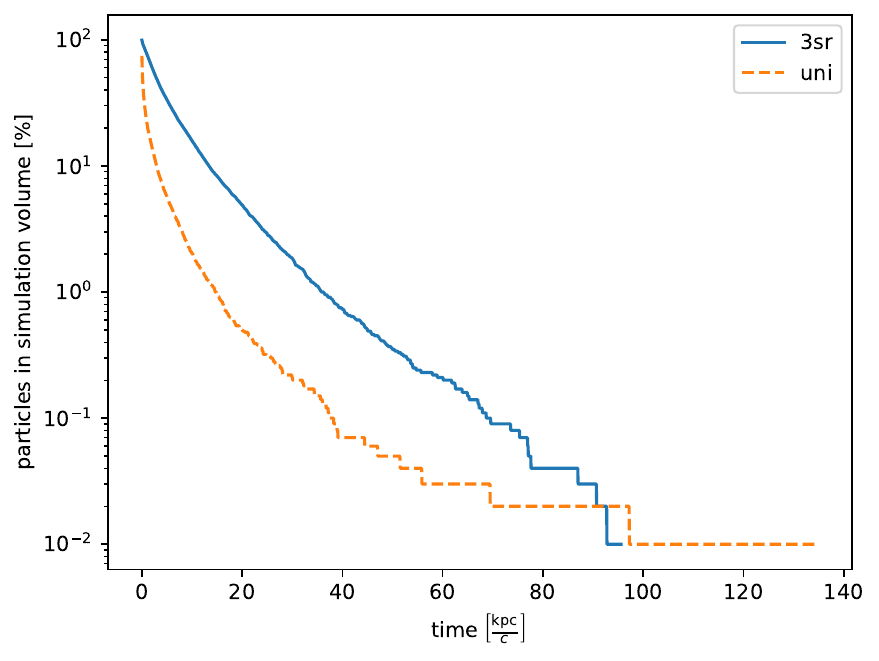}
    \caption{Number of particles left in the simulation volume.}
    \label{app:fig:particles}
\end{figure}

In general the escape time is shorter for a less anisotropic diffusion. 
In Fig.~\ref{app:fig:particles} the fraction of particles in the simulation volume is shown as a function of time. Here the cases with $\epsilon = 10^{-3}$ are chosen, to show the longest residence time. In both cases more than 99\% of the particles left the volume before $t = 100 \frac{\kpc}{c}$. Therefore the total simulation time with $T_\mathrm{max} = 500 \frac{\kpc}{c}$ is even more conservative. 

\section{Raw data for 2D countmaps} \label{app:sec:countmap}
\begin{figure*}[hp]
    \centering
    \includegraphics[height=.95\textheight]{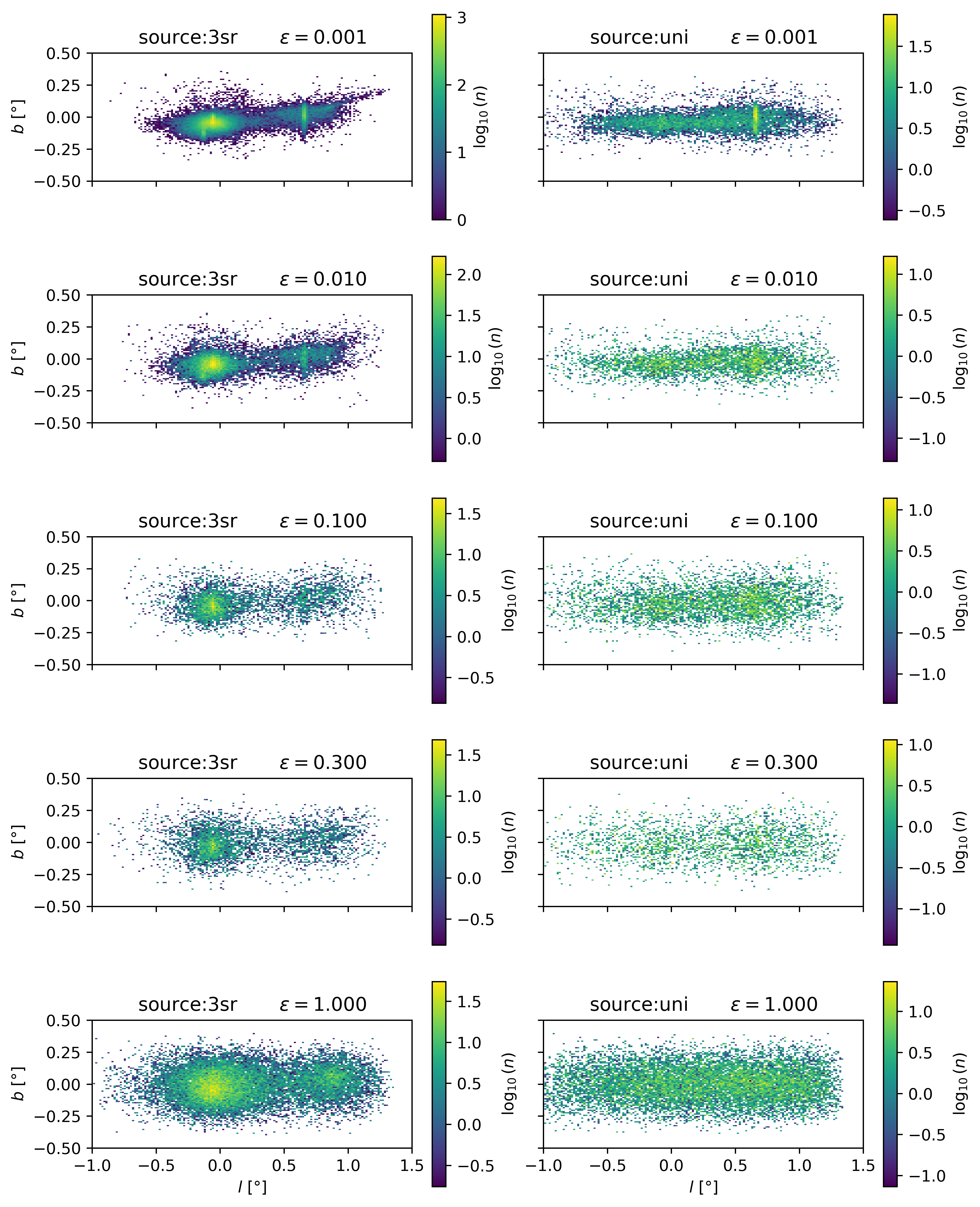}
    \caption{Synthetic $\gamma$-ray maps as shown in Fig.~\ref{fig:2d_countmaps}, but without a smearing for a PSF.}
    \label{app:fig:2d_countmap}
\end{figure*}

In figure \ref{app:fig:2d_countmap} the raw data for the synthetic $\gamma$-ray count maps are shown. The underlying binning uses $\Delta l = 0.016^\circ$ and $\Delta b = 0.01^\circ$. The simulation allows in principle for a finer binning but the statistics in each bin decreases. The bin size is chosen to minimize the noise, but keep the small-scale structures visible. 

In the case of strong parallel transport, the small-scale structures of the magnetic field can be seen. For both source distributions the impact of the NTF called \textit{radio arc} can bee seen. In the case of the \psr source distribution also smaller filaments in the region around \sgrA are visible. 

For higher values of the isotropy parameter $\epsilon$ this effect is smeared out. For $\epsilon = 0.01$ only the \textit{radio arc} is visible for the \psr distribution by eye. In the other cases no small-scale structure appears. 

\section{Impact of source spectra on count profile} \label{app:sec:source_spectrum}

\begin{figure*}[htb]
    \centering
    \includegraphics[width=.7\textwidth]{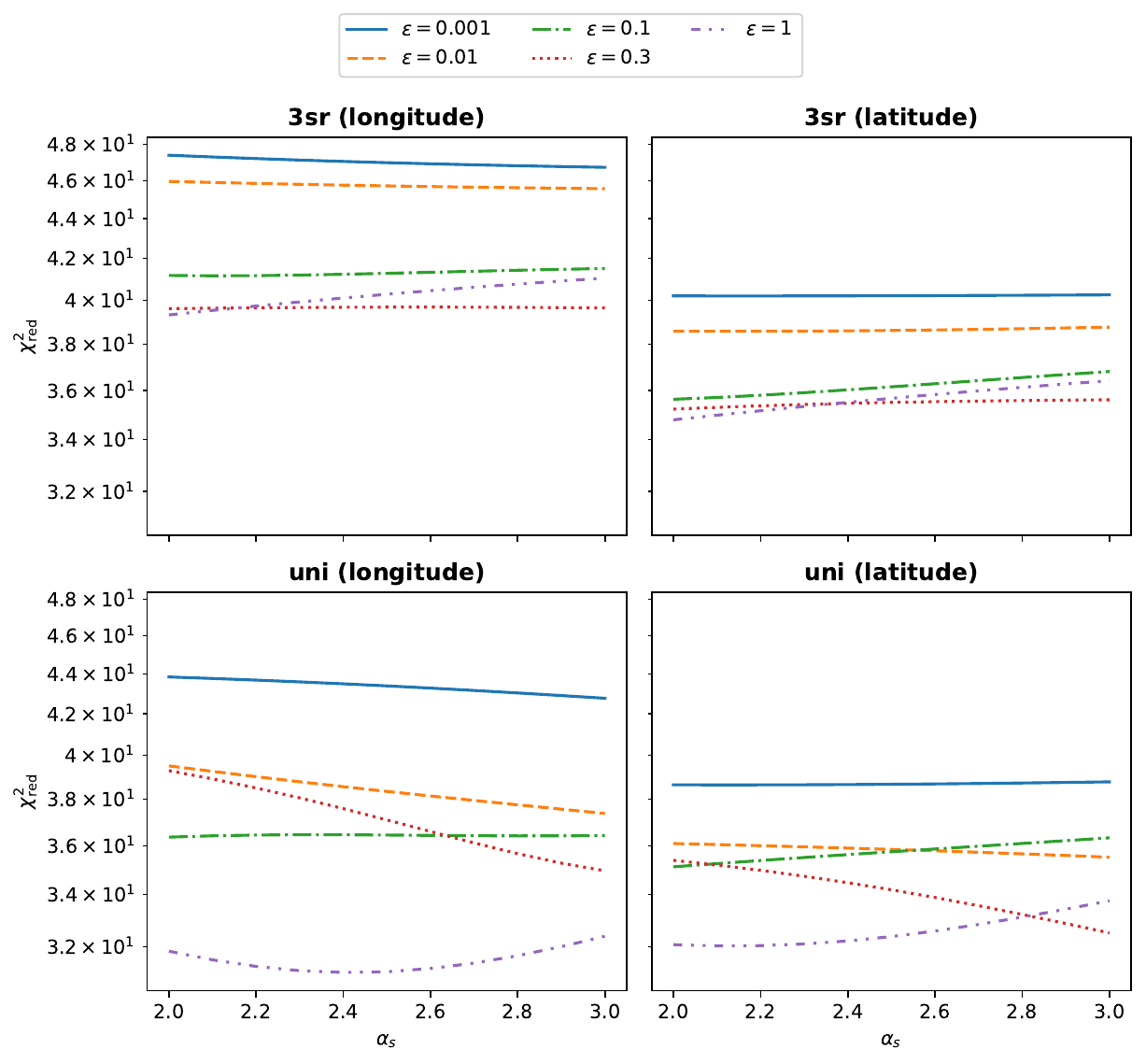}
    \caption{Impact of the source power law index $\alpha_s$ on the reduced $\chi^2$ difference between the simulation and the observed \HESS profile.}
    \label{app:fig:chi2_alpha}
\end{figure*}

In section \ref{ssec:profiles} the impact of the anisotropy on the resulting count profiles is shown. This analysis was only done for the case of a power-law injection with a slope $\alpha_s = 2$, which is consistent with the results from section \ref{ssec:SED}. In general no strong trend is observed. Only for the most extreme source indices $\alpha_s \geq 2.7$ some changes in the preferred anisotropy can be seen. But this steep injection scenario can be excluded from SED fitting in section \ref{ssec:SED}.

The general statement in sec.~\ref{ssec:profiles} does not change due to the choice of $\alpha_s = 2.0$.

%% file: main.bbl
\begin{thebibliography}{}
\expandafter\ifx\csname natexlab\endcsname\relax\def\natexlab#1{#1}\fi
\providecommand{\url}[1]{\href{#1}{#1}}
\providecommand{\dodoi}[1]{doi:~\href{http://doi.org/#1}{\nolinkurl{#1}}}
\providecommand{\doeprint}[1]{\href{http://ascl.net/#1}{\nolinkurl{http://ascl.net/#1}}}
\providecommand{\doarXiv}[1]{\href{https://arxiv.org/abs/#1}{\nolinkurl{https://arxiv.org/abs/#1}}}

\bibitem[{{Abbasi} {et~al.}(2023){Abbasi}, {Ackermann}, {Adams}, {Agarwalla},
  {Aguilar}, {Ahlers}, {Alameddine}, {Amin}, {Andeen}, {Anton}, \&
  et~al.}]{IceCube2023}
{Abbasi}, R., {Ackermann}, M., {Adams}, J., {et~al.} 2023, \apj, 956, 20,
  \dodoi{10.3847/1538-4357/acf713}

\bibitem[{Abdalla {et~al.}(2018)Abdalla, Abramowski, Aharonian, Benkhali,
  Akhperjanian, Andersson, Angüner, Arakawa, Arrieta, Aubert, Backes, Balzer,
  Barnard, Becherini, Tjus, Berge, Bernhard, Bernlöhr, Blackwell, Böttcher,
  Boisson, Bolmont, Bonnefoy, Bordas, Bregeon, Brun, Brun, Bryan, Büchele,
  Bulik, Capasso, Carr, Casanova, Cerruti, Chakraborty, Chaves, Chen,
  Chevalier, Coffaro, Colafrancesco, Cologna, Condon, Conrad, Cui, Davids,
  Decock, Degrange, Deil, Devin, DeWilt, Dirson, Djannati-Ataï, Domainko,
  Donath, Drury, Dutson, Dyks, Edwards, Egberts, Eger, Ernenwein, Eschbach,
  Farnier, Fegan, Fernandes, Fiasson, Fontaine, Förster, Funk, Füßling,
  Gabici, Gallant, Garrigoux, Giavitto, Giebels, Glicenstein, Gottschall,
  Goyal, Grondin, Hahn, Haupt, Hawkes, Heinzelmann, Henri, Hermann, Hinton,
  Hofmann, Hoischen, Holch, Holler, Horns, Ivascenko, Iwasaki, Jacholkowska,
  Jamrozy, Janiak, Jankowsky, Jankowsky, Jingo, Jogler, Jouvin, Jung-Richardt,
  Kastendieck, Katarzyński, Katsuragawa, Katz, Kerszberg, Khangulyan,
  Khélifi, King, Klepser, Klochkov, Kluźniak, Kolitzus, Komin, Kosack,
  Krakau, Kraus, Krüger, Laffon, Lamanna, Lau, Lees, Lefaucheur, Lefranc,
  Lemière, Lemoine-Goumard, Lenain, Leser, Lohse, Lorentz, Liu, López-Coto,
  Lypova, Marandon, Marcowith, Mariaud, Marx, Maurin, Maxted, Mayer, Meintjes,
  Meyer, Mitchell, Moderski, Mohamed, Mohrmann, Morå, Moulin, Murach,
  Nakashima, de~Naurois, Niederwanger, Niemiec, Oakes, O’Brien, Odaka, Ohm,
  Ostrowski, Oya, Padovani, Panter, Parsons, Pekeur, Pelletier, Perennes,
  Petrucci, Peyaud, Piel, Pita, Poon, Prokhorov, Prokoph, Pühlhofer, Punch,
  Quirrenbach, Raab, Rauth, Reimer, Reimer, Renaud, de~los Reyes, Richter,
  Rieger, Romoli, Rowell, Rudak, Rulten, Sahakian, Saito, Salek, Sanchez,
  Santangelo, Sasaki, Schlickeiser, Schüssler, Schulz, Schwanke, Schwemmer,
  Seglar-Arroyo, Settimo, Seyffert, Shafi, Shilon, Simoni, Sol, Spanier,
  Spengler, Spies, Stawarz, Steenkamp, Stegmann, Stycz, Sushch, Takahashi,
  Tavernet, Tavernier, Taylor, Terrier, Tibaldo, Tiziani, Tluczykont, Trichard,
  Tsuji, Tuffs, Uchiyama, van~der Walt, van Eldik, van Rensburg, van Soelen,
  Vasileiadis, Veh, Venter, Viana, Vincent, Vink, Voisin, Völk, Vuillaume,
  Wadiasingh, Wagner, Wagner, Wagner, White, Wierzcholska, Willmann, Wörnlein,
  Wouters, Yang, Zaborov, Zacharias, Zanin, Zdziarski, Zech, Zefi, Ziegler, \&
  Żywucka}]{HESS18}
Abdalla, H., Abramowski, A., Aharonian, F., {et~al.} 2018, \aap, 612, A9,
  \dodoi{10.1051/0004-6361/201730824}

\bibitem[{Abramowski {et~al.}(2016)Abramowski, Aharonian, Benkhali,
  Akhperjanian, Anguner, Backes, Balzer, Becherini, Tjus, Berge, Bernhard,
  Bernlohr, Birsin, Blackwell, Bottcher, Boisson, Bolmont, Bordas, Bregeon,
  Brun, Brun, Bryan, Bulik, Carr, Casanova, Chakraborty, Chalme-Calvet, Chaves,
  Chen, Chretien, Colafrancesco, Cologna, Conrad, Couturier, Cui, Davids,
  Degrange, Deil, DeWilt, Djannati-Atai, Domainko, Donath, Drury, Dubus,
  Dutson, Dyks, Dyrda, Edwards, Egberts, Eger, Ernenwein, Espigat, Farnier,
  Fegan, Feinstein, Fernandes, Fernandez, Fiasson, Fontaine, Forster, Fusling,
  Gabici, Gajdus, Gallant, Garrigoux, Giavitto, Giebels, Glicenstein,
  Gottschall, Goyal, Grondin, Grudzińska, Hadasch, Haffner, Hahn, Hawkes,
  Heinzelmann, Henri, Hermann, Hervet, Hillert, Hinton, Hofmann, Hofverberg,
  Hoischen, Holler, Horns, Ivascenko, Jacholkowska, Jamrozy, Janiak, Jankowsky,
  Jung-Richardt, Kastendieck, Katarzyński, Katz, Kerszberg, Khelifi, Kieffer,
  Klepser, Klochkov, Kluźniak, Kolitzus, Komin, Kosack, Krakau, Krayzel,
  Kruger, Laffon, Lamanna, Lau, Lefaucheur, Lefranc, Lemiere, Lemoine-Goumard,
  Lenain, Lohse, Lopatin, Lu, Lui, Marandon, Marcowith, Mariaud, Marx, Maurin,
  Maxted, Mayer, Meintjes, Menzler, Meyer, Mitchell, Moderski, Mohamed, Mora,
  Moulin, Murach, Naurois, Niemiec, Oakes, Odaka, Ottl, Ohm, Opitz, Ostrowski,
  Oya, Panter, Parsons, Arribas, Pekeur, Pelletier, Petrucci, Peyaud, Pita,
  Poon, Prokoph, Puhlhofer, Punch, Quirrenbach, Raab, Reichardt, Reimer,
  Reimer, Renaud, Reyes, Rieger, Romoli, Rosier-Lees, Rowell, Rudak, Rulten,
  Sahakian, Salek, Sanchez, Santangelo, Sasaki, Schlickeiser, Schussler,
  Schulz, Schwanke, Schwemmer, Seyffert, Simoni, Sol, Spanier, Spengler, Spies,
  Stawarz, Steenkamp, Stegmann, Stinzing, Stycz, Sushch, Tavernet, Tavernier,
  Taylor, Terrier, Tluczykont, Trichard, Tuffs, Valerius, Walt, Eldik, Soelen,
  Vasileiadis, Veh, Venter, Viana, Vincent, Vink, Voisin, Volk, Vuillaume,
  Wagner, Wagner, Wagner, Weidinger, Weitzel, White, Wierzcholska, Willmann,
  Wornlein, Wouters, Yang, Zabalza, Zaborov, Zacharias, Zdziarski, Zech, Zefi,
  \& Zywucka}]{HESS16}
Abramowski, A., Aharonian, F., Benkhali, F.~A., {et~al.} 2016, Nature, 531,
  476, \dodoi{10.1038/nature17147}

\bibitem[{Acciari {et~al.}(2020)Acciari, Ansoldi, Antonelli, Engels, Baack,
  Babić, Banerjee, de~Almeida, Barrio, González, Bednarek, Bellizzi,
  Bernardini, Berti, Besenrieder, Bhattacharyya, Bigongiari, Biland, Blanch,
  Bonnoli, Bošnjak, Busetto, Carosi, Ceribella, Chai, Chilingaryan, Cikota,
  Colak, Colin, Colombo, Contreras, Cortina, Covino, D’Elia, Vela, Dazzi,
  Angelis, Lotto, Delfino, Delgado, Depaoli, Pierro, Venere, Espiñeira,
  Prester, Donini, Dorner, Doro, Elsaesser, Ramazani, Fattorini,
  Fernández-Barral, Ferrara, Fidalgo, Foffano, Fonseca, Font, Fruck, Fukami,
  López, Garczarczyk, Gasparyan, Gaug, Giglietto, Giordano, Godinović, Green,
  Guberman, Hadasch, Hahn, Herrera, Hoang, Hrupec, Hütten, Inada, Inoue,
  Ishio, Iwamura, Jouvin, Kerszberg, Kubo, Kushida, Lamastra, Lelas, Leone,
  Lindfors, Lombardi, Longo, López, López-Coto, López-Oramas, Loporchio,
  de~Oliveira~Fraga, Maggio, Majumdar, Makariev, Mallamaci, Maneva, Manganaro,
  Mannheim, Maraschi, Mariotti, Martínez, Masuda, Mazin, Mićanović, Miceli,
  Minev, Miranda, Mirzoyan, Molina, Moralejo, Morcuende, Moreno, Moretti,
  Munar-Adrover, Neustroev, Nigro, Nilsson, Ninci, Nishijima, Noda, Nogués,
  Nöthe, Nozaki, Paiano, Palacio, Palatiello, Paneque, Paoletti, Paredes,
  Peñil, Peresano, Persic, Moroni, Prandini, Puljak, Rhode, Ribó, Rico,
  Righi, Rugliancich, Saha, Sahakyan, Saito, Sakurai, Satalecka, Schmidt,
  Schweizer, Sitarek, Šnidarić, Sobczynska, Somero, Stamerra, Strom, Strzys,
  Suda, Surić, Takahashi, Tavecchio, Temnikov, Terzić, Teshima, Torres-Albà,
  Tosti, Tsujimoto, Vagelli, van Scherpenberg, Vanzo, Acosta, Vigorito, Vitale,
  Vovk, Will, \& Zarić}]{Magic20}
Acciari, V.~A., Ansoldi, S., Antonelli, L.~A., {et~al.} 2020, \aap, 642, A190,
  \dodoi{10.1051/0004-6361/201936896}

\bibitem[{{Ackermann} {et~al.}(2014){Ackermann}, {Albert}, {Atwood}, {Baldini},
  {Ballet}, {Barbiellini}, {Bastieri}, {Bellazzini}, {Bissaldi}, {Blandford},
  {Bloom}, {Bottacini}, {Brandt}, {Bregeon}, {Bruel}, {Buehler}, {Buson},
  {Caliandro}, {Cameron}, {Caragiulo}, {Caraveo}, {Cavazzuti}, {Cecchi},
  {Charles}, {Chekhtman}, {Chiang}, {Chiaro}, {Ciprini}, {Claus},
  {Cohen-Tanugi}, {Conrad}, {Cutini}, {D'Ammando}, {de Angelis}, {de Palma},
  {Dermer}, {Digel}, {Di Venere}, {Silva}, {Drell}, {Favuzzi}, {Ferrara},
  {Focke}, {Franckowiak}, {Fukazawa}, {Funk}, {Fusco}, {Gargano}, {Gasparrini},
  {Germani}, {Giglietto}, {Giordano}, {Giroletti}, {Godfrey}, {Gomez-Vargas},
  {Grenier}, {Guiriec}, {Hadasch}, {Harding}, {Hays}, {Hewitt}, {Hou},
  {Jogler}, {J{\'o}hannesson}, {Johnson}, {Johnson}, {Kamae}, {Kataoka},
  {Kn{\"o}dlseder}, {Kocevski}, {Kuss}, {Larsson}, {Latronico}, {Longo},
  {Loparco}, {Lovellette}, {Lubrano}, {Malyshev}, {Manfreda}, {Massaro},
  {Mayer}, {Mazziotta}, {McEnery}, {Michelson}, {Mitthumsiri}, {Mizuno},
  {Monzani}, {Morselli}, {Moskalenko}, {Murgia}, {Nemmen}, {Nuss}, {Ohsugi},
  {Omodei}, {Orienti}, {Orlando}, {Ormes}, {Paneque}, {Panetta}, {Perkins},
  {Pesce-Rollins}, {Petrosian}, {Piron}, {Pivato}, {Rain{\`o}}, {Rando},
  {Razzano}, {Razzaque}, {Reimer}, {Reimer}, {S{\'a}nchez-Conde}, {Schaal},
  {Schulz}, {Sgr{\`o}}, {Siskind}, {Spandre}, {Spinelli}, {Stawarz}, {Strong},
  {Suson}, {Tahara}, {Takahashi}, {Thayer}, {Tibaldo}, {Tinivella}, {Torres},
  {Tosti}, {Troja}, {Uchiyama}, {Vianello}, {Werner}, {Winer}, {Wood}, {Wood},
  \& {Zaharijas}}]{Ackermann14}
{Ackermann}, M., {Albert}, A., {Atwood}, W.~B., {et~al.} 2014, \apj, 793, 64,
  \dodoi{10.1088/0004-637X/793/1/64}

\bibitem[{Ackermann {et~al.}(2017)Ackermann, Ajello, Albert, Atwood, Baldini,
  Ballet, Barbiellini, Bastieri, Bellazzini, Bissaldi, Blandford, Bloom,
  Bonino, Bottacini, Brandt, Bregeon, Bruel, Buehler, Burnett, Cameron, Caputo,
  Caragiulo, Caraveo, Cavazzuti, Cecchi, Charles, Chekhtman, Chiang, Chiappo,
  Chiaro, Ciprini, Conrad, Costanza, Cuoco, Cutini, D’Ammando, de~Palma,
  Desiante, Digel, Lalla, Mauro, Venere, Drell, Favuzzi, Fegan, Ferrara, Focke,
  Franckowiak, Fukazawa, Funk, Fusco, Gargano, Gasparrini, Giglietto, Giordano,
  Giroletti, Glanzman, Gomez-Vargas, Green, Grenier, Grove, Guillemot, Guiriec,
  Gustafsson, Harding, Hays, Hewitt, Horan, Jogler, Johnson, Kamae, Kocevski,
  Kuss, Mura, Larsson, Latronico, Li, Longo, Loparco, Lovellette, Lubrano,
  Magill, Maldera, Malyshev, Manfreda, Martin, Mazziotta, Michelson, Mirabal,
  Mitthumsiri, Mizuno, Moiseev, Monzani, Morselli, Negro, Nuss, Ohsugi,
  Orienti, Orlando, Ormes, Paneque, Perkins, Persic, Pesce-Rollins, Piron,
  Principe, Rainò, Rando, Razzano, Razzaque, Reimer, Reimer, Sánchez-Conde,
  Sgrò, Simone, Siskind, Spada, Spandre, Spinelli, Suson, Tajima, Tanaka,
  Thayer, Tibaldo, Torres, Troja, Uchiyama, Vianello, Wood, Wood, Zaharijas, \&
  Zimmer}]{Ackermann2017}
Ackermann, M., Ajello, M., Albert, A., {et~al.} 2017, ApJ, 840, 43,
  \dodoi{10.3847/1538-4357/aa6cab}

\bibitem[{Adams {et~al.}(2021)Adams, Benbow, Brill, Brose, Buchovecky, Capasso,
  Christiansen, Chromey, Daniel, Errando, Falcone, Feng, Finley, Fortson,
  Furniss, Gent, Gillanders, Giuri, Hanna, Hervet, Holder, Hughes, Humensky,
  Jin, Kaaret, Kelley-Hoskins, Kertzman, Kieda, Krennrich, Kumar, Lang, Lundy,
  Maier, Moriarty, Mukherjee, Nieto, Nievas-Rosillo, O'brien, Ong, Otte,
  Pfrang, Pohl, Prado, Pueschel, Quinn, Ragan, Reynolds, Ribeiro, Richards,
  Roache, Ryan, Santander, Schlenstedt, Sembroski, Shang, Stevenson, Wakely,
  Weinstein, \& Williams}]{Veritas21}
Adams, C.~B., Benbow, W., Brill, A., {et~al.} 2021, ApJ, 913, 115,
  \dodoi{10.3847/1538-4357/abf926}

\bibitem[{Ajello {et~al.}(2016)Ajello, Albert, Atwood, Barbiellini, Bastieri,
  Bechtol, Bellazzini, Bissaldi, Blandford, Bloom, Bonino, Bottacini, Brandt,
  Bregeon, Bruel, Buehler, Buson, Caliandro, Cameron, Caputo, Caragiulo,
  Caraveo, Cecchi, Chekhtman, Chiang, Chiaro, Ciprini, Cohen-Tanugi, Cominsky,
  Conrad, Cutini, D’Ammando, de~Angelis, de~Palma, Desiante, Venere, Drell,
  Favuzzi, Ferrara, Fusco, Gargano, Gasparrini, Giglietto, Giommi, Giordano,
  Giroletti, Glanzman, Godfrey, Gomez-Vargas, Grenier, Guiriec, Gustafsson,
  Harding, Hewitt, Hill, Horan, Jogler, Jóhannesson, Johnson, Kamae, Karwin,
  Knödlseder, Kuss, Larsson, Latronico, Li, Li, Longo, Loparco, Lovellette,
  Lubrano, Magill, Maldera, Malyshev, Manfreda, Mayer, Mazziotta, Michelson,
  Mitthumsiri, Mizuno, Moiseev, Monzani, Morselli, Moskalenko, Murgia, Nuss,
  Ohno, Ohsugi, Omodei, Orlando, Ormes, Paneque, Pesce-Rollins, Piron, Pivato,
  Porter, Rainò, Rando, Razzano, Reimer, Reimer, Ritz, Sánchez-Conde,
  Parkinson, Sgrò, Siskind, Smith, Spada, Spandre, Spinelli, Suson, Tajima,
  Takahashi, Thayer, Torres, Tosti, Troja, Uchiyama, Vianello, Winer, Wood,
  Zaharijas, \& Zimmer}]{Ajello2016}
Ajello, M., Albert, A., Atwood, W.~B., {et~al.} 2016, ApJ, 819, 44,
  \dodoi{10.3847/0004-637X/819/1/44}

\bibitem[{Batista {et~al.}(2016)Batista, Dundovic, Erdmann, Kampert, Kuempel,
  Müller, Sigl, Vliet, Walz, \& Winchen}]{CRPropa3}
Batista, R.~A., Dundovic, A., Erdmann, M., {et~al.} 2016, \jcap, 2016, 038,
  \dodoi{10.1088/1475-7516/2016/05/038}

\bibitem[{Batista {et~al.}(2022)Batista, Tjus, Dörner, Dundovic, Eichmann,
  Frie, Heiter, Hoerbe, Kampert, Merten, Müller, Reichherzer, Saveliev,
  Schlegel, Sigl, van Vliet, \& Winchen}]{CRPropa32}
Batista, R.~A., Tjus, J.~B., Dörner, J., {et~al.} 2022, \jcap, 2022, 035,
  \dodoi{10.1088/1475-7516/2022/09/035}

\bibitem[{{Becker Tjus} \& {Merten}(2020)}]{Tjus2020}
{Becker Tjus}, J., \& {Merten}, L. 2020, \physrep, 872, 1,
  \dodoi{10.1016/j.physrep.2020.05.002}

\bibitem[{Cao {et~al.}(2023)Cao, Aharonian, An, Axikegu, Bai, Bao, Bastieri,
  Bi, Bi, Cai, Cao, Cao, Cao, Chang, Chang, Chen, Chen, Chen, Chen, Chen, Chen,
  Chen, Chen, Chen, Chen, Chen, Chen, Cheng, Cheng, Cui, Cui, Cui, Cui, Dai,
  Dai, Dai, Danzengluobu, della Volpe, Dong, Duan, Fan, Fan, Fang, Fang, Feng,
  Feng, Feng, Feng, Feng, Gabici, Gao, Gao, Gao, Gao, Gao, Gao, Ge, Geng,
  Giacinti, Gong, Gou, Gu, Guo, Guo, Guo, Guo, Han, He, He, He, He, He, Heller,
  Hor, Hou, Hou, Hou, Hu, Hu, Hu, Huang, Huang, Huang, Huang, Huang, Huang,
  Huang, Ji, Jia, Jia, Jiang, Jiang, Jiang, Jin, Kang, Ke, Kuleshov, Kurinov,
  Li, Li, Li, Li, Li, Li, Li, Li, Li, Li, Li, Li, Li, Li, Li, Li, Li, Li, Li,
  Liang, Liang, Lin, Liu, Liu, Liu, Liu, Liu, Liu, Liu, Liu, Liu, Liu, Liu,
  Liu, Liu, Liu, Lu, Luo, Lv, Ma, Ma, Ma, Mao, Min, Mitthumsiri, Mu, Nan,
  Neronov, Ou, Pang, Pattarakijwanich, Pei, Qi, Qi, Qiao, Qin, Ruffolo, Sáiz,
  Semikoz, Shao, Shao, Shchegolev, Sheng, Shu, Song, Stenkin, Stepanov, Su,
  Sun, Sun, Sun, Tam, Tang, Tang, Tian, Wang, Wang, Wang, Wang, Wang, Wang,
  Wang, Wang, Wang, Wang, Wang, Wang, Wang, Wang, Wang, Wang, Wang, Wang, Wang,
  Wang, Wang, Wei, Wei, Wei, Wen, Wu, Wu, Wu, Wu, Wu, Xi, Xia, Xia, Xiang,
  Xiao, Xiao, Xin, Xin, Xing, Xiong, Xu, Xu, Xu, Xu, Xue, Yan, Yan, Yan, Yang,
  Yang, Yang, Yang, Yang, Yang, Yang, Yang, Yang, Yao, Yao, Ye, Yin, Yin, You,
  You, Yu, Yuan, Yue, Zeng, Zeng, Zeng, Zha, Zhang, Zhang, Zhang, Zhang, Zhang,
  Zhang, Zhang, Zhang, Zhang, Zhang, Zhang, Zhang, Zhang, Zhang, Zhang, Zhang,
  Zhang, Zhang, Zhao, Zhao, Zhao, Zhao, Zhao, Zheng, Zhou, Zhou, Zhou, Zhou,
  Zhou, Zhou, Zhou, Zhu, Zhu, Zhu, Zhu, \& Zuo}]{Cao2023}
Cao, Z., Aharonian, F., An, Q., {et~al.} 2023, Physical Review Letters, 131,
  151001, \dodoi{10.1103/PhysRevLett.131.151001}

\bibitem[{{Cerri} {et~al.}(2017){Cerri}, {Gaggero}, {Vittino}, {Evoli}, \&
  {Grasso}}]{Cerri-etal-2017}
{Cerri}, S.~S., {Gaggero}, D., {Vittino}, A., {Evoli}, C., \& {Grasso}, D.
  2017, \jcap, 2017, 019, \dodoi{10.1088/1475-7516/2017/10/019}

\bibitem[{{Daylan} {et~al.}(2016){Daylan}, {Finkbeiner}, {Hooper}, {Linden},
  {Portillo}, {Rodd}, \& {Slatyer}}]{Daylan2016}
{Daylan}, T., {Finkbeiner}, D.~P., {Hooper}, D., {et~al.} 2016, Physics of the
  Dark Universe, 12, 1, \dodoi{10.1016/j.dark.2015.12.005}

\bibitem[{Di~Mauro(2021)}]{DiMauro:2021raz}
Di~Mauro, M. 2021, Phys. Rev. D, 103, 063029,
  \dodoi{10.1103/PhysRevD.103.063029}

\bibitem[{{Effenberger} {et~al.}(2012{\natexlab{a}}){Effenberger}, {Fichtner},
  {Scherer}, {Barra}, {Kleimann}, \& {Strauss}}]{Effenberger-etal-2012}
{Effenberger}, F., {Fichtner}, H., {Scherer}, K., {et~al.} 2012{\natexlab{a}},
  \apj, 750, 108, \dodoi{10.1088/0004-637X/750/2/108}

\bibitem[{{Effenberger} {et~al.}(2012{\natexlab{b}}){Effenberger}, {Fichtner},
  {Scherer}, \& {B{\"u}sching}}]{Effenberger-etal-2012b}
{Effenberger}, F., {Fichtner}, H., {Scherer}, K., \& {B{\"u}sching}, I.
  2012{\natexlab{b}}, \aap, 547, A120, \dodoi{10.1051/0004-6361/201220203}

\bibitem[{Eichmann \& Kachelrieß(2023)}]{Eichmann2023}
Eichmann, B., \& Kachelrieß, M. 2023, Journal of Cosmology and Astroparticle
  Physics, 2023, 053, \dodoi{10.1088/1475-7516/2023/02/053}

\bibitem[{{Ferri\`ere} {et~al.}(2007){Ferri\`ere}, {Gillard}, \&
  {Jean}}]{Ferriere07}
{Ferri\`ere}, K., {Gillard}, W., \& {Jean}, P. 2007, A\&A, 467, 611,
  \dodoi{10.1051/0004-6361:20066992}

\bibitem[{{Finkbeiner}(2004)}]{Finkbeiner04}
{Finkbeiner}, D.~P. 2004, \apj, 614, 186, \dodoi{10.1086/423482}

\bibitem[{Goodenough \& Hooper(2009)}]{Goodenough2009}
Goodenough, L., \& Hooper, D. 2009, arXiv:0910.2998

\bibitem[{{Guenduez} {et~al.}(2020){Guenduez}, {Becker Tjus}, {Ferri{\`e}re},
  \& {Dettmar}}]{guenduez_bfield2020}
{Guenduez}, M., {Becker Tjus}, J., {Ferri{\`e}re}, K., \& {Dettmar}, R.~J.
  2020, \aap, 644, A71, \dodoi{10.1051/0004-6361/201936081}

\bibitem[{{H.~E.~S.~S. Collaboration} {et~al.}(2018){H.~E.~S.~S.
  Collaboration}, {Abdalla}, {Abramowski}, {Aharonian}, {Ait Benkhali},
  {Ang{\"u}ner}, {Arakawa}, {Arrieta}, {Aubert}, {Backes}, {Balzer}, {Barnard},
  {Becherini}, {Becker Tjus}, {Berge}, {Bernhard}, {Bernl{\"o}hr}, {Blackwell},
  {B{\"o}ttcher}, {Boisson}, {Bolmont}, {Bonnefoy}, {Bordas}, {Bregeon},
  {Brun}, {Brun}, {Bryan}, {B{\"u}chele}, {Bulik}, {Capasso}, {Carrigan},
  {Caroff}, {Carosi}, {Casanova}, {Cerruti}, {Chakraborty}, {Chaves}, {Chen},
  {Chevalier}, {Colafrancesco}, {Condon}, {Conrad}, {Davids}, {Decock}, {Deil},
  {Devin}, {deWilt}, {Dirson}, {Djannati-Ata{\"\i}}, {Domainko}, {Donath},
  {Drury}, {Dutson}, {Dyks}, {Edwards}, {Egberts}, {Eger}, {Emery},
  {Ernenwein}, {Eschbach}, {Farnier}, {Fegan}, {Fernandes}, {Fiasson},
  {Fontaine}, {F{\"o}rster}, {Funk}, {F{\"u}{\ss}ling}, {Gabici}, {Gallant},
  {Garrigoux}, {Gast}, {Gat{\'e}}, {Giavitto}, {Giebels}, {Glawion},
  {Glicenstein}, {Gottschall}, {Grondin}, {Hahn}, {Haupt}, {Hawkes},
  {Heinzelmann}, {Henri}, {Hermann}, {Hinton}, {Hofmann}, {Hoischen}, {Holch},
  {Holler}, {Horns}, {Ivascenko}, {Iwasaki}, {Jacholkowska}, {Jamrozy},
  {Jankowsky}, {Jankowsky}, {Jingo}, {Jouvin}, {Jung-Richardt}, {Kastendieck},
  {Katarzy{\'n}ski}, {Katsuragawa}, {Katz}, {Kerszberg}, {Khangulyan},
  {Kh{\'e}lifi}, {King}, {Klepser}, {Klochkov}, {Klu{\'z}niak}, {Komin},
  {Kosack}, {Krakau}, {Kraus}, {Kr{\"u}ger}, {Laffon}, {Lamanna}, {Lau},
  {Lees}, {Lefaucheur}, {Lemi{\`e}re}, {Lemoine-Goumard}, {Lenain}, {Leser},
  {Lohse}, {Lorentz}, {Liu}, {L{\'o}pez-Coto}, {Lypova}, {Marandon},
  {Malyshev}, {Marcowith}, {Mariaud}, {Marx}, {Maurin}, {Maxted}, {Mayer},
  {Meintjes}, {Meyer}, {Mitchell}, {Moderski}, {Mohamed}, {Mohrmann},
  {Mor{\r{a}}}, {Moulin}, {Murach}, {Nakashima}, {de Naurois}, {Ndiyavala},
  {Niederwanger}, {Niemiec}, {Oakes}, {O'Brien}, {Odaka}, {Ohm}, {Ostrowski},
  {Oya}, {Padovani}, {Panter}, {Parsons}, {Paz Arribas}, {Pekeur}, {Pelletier},
  {Perennes}, {Petrucci}, {Peyaud}, {Piel}, {Pita}, {Poireau}, {Poon},
  {Prokhorov}, {Prokoph}, {P{\"u}hlhofer}, {Punch}, {Quirrenbach}, {Raab},
  {Rauth}, {Reimer}, {Reimer}, {Renaud}, {de los Reyes}, {Rieger}, {Rinchiuso},
  {Romoli}, {Rowell}, {Rudak}, {Rulten}, {Safi-Harb}, {Sahakian}, {Saito},
  {Sanchez}, {Santangelo}, {Sasaki}, {Schandri}, {Schlickeiser},
  {Sch{\"u}ssler}, {Schulz}, {Schwanke}, {Schwemmer}, {Seglar-Arroyo},
  {Settimo}, {Seyffert}, {Shafi}, {Shilon}, {Shiningayamwe}, {Simoni}, {Sol},
  {Spanier}, {Spir-Jacob}, {Stawarz}, {Steenkamp}, {Stegmann}, {Steppa},
  {Sushch}, {Takahashi}, {Tavernet}, {Tavernier}, {Taylor}, {Terrier},
  {Tibaldo}, {Tiziani}, {Tluczykont}, {Trichard}, {Tsirou}, {Tsuji}, {Tuffs},
  {Uchiyama}, {van der Walt}, {van Eldik}, {van Rensburg}, {van Soelen},
  {Vasileiadis}, {Veh}, {Venter}, {Viana}, {Vincent}, {Vink}, {Voisin},
  {V{\"o}lk}, {Vuillaume}, {Wadiasingh}, {Wagner}, {Wagner}, {Wagner}, {White},
  {Wierzcholska}, {Willmann}, {W{\"o}rnlein}, {Wouters}, {Yang}, {Zaborov},
  {Zacharias}, {Zanin}, {Zdziarski}, {Zech}, {Zefi}, {Ziegler}, {Zorn}, \&
  {{\.Z}ywucka}}]{HESS18_GPS}
{H.~E.~S.~S. Collaboration}, {Abdalla}, H., {Abramowski}, A., {et~al.} 2018,
  \aap, 612, A1, \dodoi{10.1051/0004-6361/201732098}

\bibitem[{Harris {et~al.}(2020)Harris, Millman, van~der Walt, Gommers,
  Virtanen, Cournapeau, Wieser, Taylor, Berg, Smith, Kern, Picus, Hoyer, van
  Kerkwijk, Brett, Haldane, del R{\'{i}}o, Wiebe, Peterson,
  G{\'{e}}rard-Marchant, Sheppard, Reddy, Weckesser, Abbasi, Gohlke, \&
  Oliphant}]{numpy}
Harris, C.~R., Millman, K.~J., van~der Walt, S.~J., {et~al.} 2020, Nature, 585,
  357, \dodoi{10.1038/s41586-020-2649-2}

\bibitem[{{Henshaw} {et~al.}(2023){Henshaw}, {Barnes}, {Battersby}, {Ginsburg},
  {Sormani}, \& {Walker}}]{Henshaw23}
{Henshaw}, J.~D., {Barnes}, A.~T., {Battersby}, C., {et~al.} 2023, in
  Astronomical Society of the Pacific Conference Series, Vol. 534, Protostars
  and Planets VII, 83, \dodoi{10.48550/arXiv.2203.11223}

\bibitem[{Heywood {et~al.}(2022)Heywood, Rammala, Camilo, Cotton, Yusef-Zadeh,
  Abbott, Adam, Adams, Aldera, Asad, Bauermeister, Bennett, Bester, Bode,
  Botha, Botha, Brederode, Buchner, Burger, Cheetham, de~Villiers,
  Dikgale-Mahlakoana, du~Toit, Esterhuyse, Fanaroff, February, Fourie, Frank,
  Gamatham, Geyer, Goedhart, Gouws, Gumede, Hlakola, Hokwana, Hoosen, Horrell,
  Hugo, Isaacson, Józsa, Jonas, Joubert, Julie, Kapp, Kenyon, Kotzé, Kriek,
  Kriel, Krishnan, Lehmensiek, Liebenberg, Lord, Lunsky, Madisa, Magnus,
  Mahgoub, Makhaba, Makhathini, Malan, Manley, Marais, Martens, Mauch, Merry,
  Millenaar, Mnyandu, Mokone, Monama, Mphego, New, Ngcebetsha, Ngoasheng,
  Ockards, Oozeer, Otto, Passmoor, Patel, Peens-Hough, Perkins, Ramaila,
  Ramanujam, Ramudzuli, Ratcliffe, Robyntjies, Salie, Sambu, Schollar,
  Schwardt, Schwartz, Serylak, Siebrits, Sirothia, Slabber, Smirnov, Sofeya,
  Taljaard, Tasse, Tiplady, Toruvanda, Twum, van Balla, van~der Byl, van~der
  Merwe, Tonder, Wyk, Venter, Venter, Wallace, Welz, Williams, \&
  Xaia}]{Heywood2022}
Heywood, I., Rammala, I., Camilo, F., {et~al.} 2022, ApJ, 925, 165,
  \dodoi{10.3847/1538-4357/ac449a}

\bibitem[{{Hoerbe} {et~al.}(2020){Hoerbe}, {Morris}, {Cotter}, \& {Becker
  Tjus}}]{Hoerbe}
{Hoerbe}, M.~R., {Morris}, P.~J., {Cotter}, G., \& {Becker Tjus}, J. 2020,
  \mnras, 496, 2885, \dodoi{10.1093/mnras/staa1650}

\bibitem[{Hunter(2007)}]{matplotlib}
Hunter, J.~D. 2007, Computing in Science \& Engineering, 9, 90,
  \dodoi{10.1109/MCSE.2007.55}

\bibitem[{{Jokipii}(1966)}]{Jokipii-1966}
{Jokipii}, J.~R. 1966, \apj, 146, 480, \dodoi{10.1086/148912}

\bibitem[{{Kelner} {et~al.}(2006){Kelner}, {Aharonian}, \&
  {Bugayov}}]{Kelner06}
{Kelner}, S.~R., {Aharonian}, F.~A., \& {Bugayov}, V.~V. 2006, \prd, 74,
  034018, \dodoi{10.1103/PhysRevD.74.034018}

\bibitem[{{Merten} {et~al.}(2017){Merten}, {Becker Tjus}, {Fichtner},
  {Eichmann}, \& {Sigl}}]{Merten17}
{Merten}, L., {Becker Tjus}, J., {Fichtner}, H., {Eichmann}, B., \& {Sigl}, G.
  2017, \jcap, 2017, 046, \dodoi{10.1088/1475-7516/2017/06/046}

\bibitem[{{Pedlar} {et~al.}(1989){Pedlar}, {Anantharamaiah}, {Ekers}, {Goss},
  {van Gorkom}, {Schwarz}, \& {Zhao}}]{Pedlar89}
{Pedlar}, A., {Anantharamaiah}, K.~R., {Ekers}, R.~D., {et~al.} 1989, \apj,
  342, 769, \dodoi{10.1086/167635}

\bibitem[{P\'erez \& Granger(2007)}]{ipython}
P\'erez, F., \& Granger, B.~E. 2007, Computing in Science and Engineering, 9,
  21, \dodoi{10.1109/MCSE.2007.53}

\bibitem[{{Planck Collaboration} {et~al.}(2013){Planck Collaboration}, {Ade},
  {Aghanim}, {Arnaud}, {Ashdown}, {Atrio-Barandela}, {Aumont}, {Baccigalupi},
  {Balbi}, {Banday}, {Barreiro}, {Bartlett}, {Battaner}, {Benabed},
  {Beno{\^\i}t}, {Bernard}, {Bersanelli}, {Bonaldi}, {Bond}, {Borrill},
  {Bouchet}, {Burigana}, {Cabella}, {Cardoso}, {Catalano}, {Cay{\'o}n},
  {Chary}, {Chiang}, {Christensen}, {Clements}, {Colombo}, {Coulais}, {Crill},
  {Cuttaia}, {Danese}, {D'Arcangelo}, {Davis}, {de Bernardis}, {de Gasperis},
  {de Rosa}, {de Zotti}, {Delabrouille}, {Dickinson}, {Diego}, {Dobler},
  {Dole}, {Donzelli}, {Dor{\'e}}, {D{\"o}rl}, {Douspis}, {Dupac}, {Efstathiou},
  {En{\ss}lin}, {Eriksen}, {Finelli}, {Forni}, {Frailis}, {Franceschi},
  {Galeotta}, {Ganga}, {Giard}, {Giardino}, {Gonz{\'a}lez-Nuevo}, {G{\'o}rski},
  {Gratton}, {Gregorio}, {Gruppuso}, {Hansen}, {Harrison}, {Helou},
  {Henrot-Versill{\'e}}, {Hern{\'a}ndez-Monteagudo}, {Hildebrandt}, {Hivon},
  {Hobson}, {Holmes}, {Hornstrup}, {Hovest}, {Huffenberger}, {Jaffe},
  {Jagemann}, {Jewell}, {Jones}, {Juvela}, {Keih{\"a}nen}, {Knoche}, {Knox},
  {Kunz}, {Kurki-Suonio}, {Lagache}, {L{\"a}hteenm{\"a}ki}, {Lamarre},
  {Lasenby}, {Lawrence}, {Leach}, {Leonardi}, {Lilje}, {Linden-V{\o}rnle},
  {L{\'o}pez-Caniego}, {Lubin}, {Mac{\'\i}as-P{\'e}rez}, {Maffei}, {Maino},
  {Mandolesi}, {Maris}, {Marshall}, {Martin}, {Mart{\'\i}nez-Gonz{\'a}lez},
  {Masi}, {Massardi}, {Matarrese}, {Matthai}, {Mazzotta}, {Meinhold},
  {Melchiorri}, {Mendes}, {Mennella}, {Mitra}, {Moneti}, {Montier}, {Morgante},
  {Munshi}, {Murphy}, {Naselsky}, {Natoli}, {N{\o}rgaard-Nielsen}, {Noviello},
  {Novikov}, {Novikov}, {Osborne}, {Pajot}, {Paladini}, {Paoletti},
  {Partridge}, {Pearson}, {Perdereau}, {Perrotta}, {Piacentini}, {Piat},
  {Pierpaoli}, {Pietrobon}, {Plaszczynski}, {Pointecouteau}, {Polenta},
  {Ponthieu}, {Popa}, {Poutanen}, {Pratt}, {Prunet}, {Puget}, {Rachen},
  {Rebolo}, {Reinecke}, {Renault}, {Ricciardi}, {Riller}, {Ristorcelli},
  {Rocha}, {Rosset}, {Rubi{\~n}o-Mart{\'\i}n}, {Rusholme}, {Sandri}, {Savini},
  {Schaefer}, {Scott}, {Smoot}, {Spencer}, {Stivoli}, {Sudiwala}, {Suur-Uski},
  {Sygnet}, {Tauber}, {Terenzi}, {Toffolatti}, {Tomasi}, {Tristram},
  {T{\"u}rler}, {Umana}, {Valenziano}, {Van Tent}, {Vielva}, {Villa},
  {Vittorio}, {Wade}, {Wandelt}, {White}, {Yvon}, {Zacchei}, \&
  {Zonca}}]{Planck13}
{Planck Collaboration}, {Ade}, P.~A.~R., {Aghanim}, N., {et~al.} 2013, \aap,
  554, A139, \dodoi{10.1051/0004-6361/201220271}

\bibitem[{Reichherzer {et~al.}(2020)Reichherzer, {Becker Tjus}, Zweibel,
  Merten, \& Pueschel}]{Reichherzer2020}
Reichherzer, P., {Becker Tjus}, J., Zweibel, E.~G., Merten, L., \& Pueschel,
  M.~J. 2020, \mnras, 498, 5051, \dodoi{10.1093/mnras/staa2533}

\bibitem[{{Reichherzer} {et~al.}(2022{\natexlab{a}}){Reichherzer}, {Becker
  Tjus}, {Zweibel}, {Merten}, \& {Pueschel}}]{Reichherzer2021b}
{Reichherzer}, P., {Becker Tjus}, J., {Zweibel}, E.~G., {Merten}, L., \&
  {Pueschel}, M.~J. 2022{\natexlab{a}}, \mnras, 514, 2658,
  \dodoi{10.1093/mnras/stac1408}

\bibitem[{{Reichherzer} {et~al.}(2022{\natexlab{b}}){Reichherzer}, {Merten},
  {D{\"o}rner}, {Becker Tjus}, {Pueschel}, \& {Zweibel}}]{Reichherzer2021}
{Reichherzer}, P., {Merten}, L., {D{\"o}rner}, J., {et~al.} 2022{\natexlab{b}},
  SN Applied Sciences, 4, 15, \dodoi{10.1007/s42452-021-04891-z}

\bibitem[{Rocklin(2015)}]{dask}
Rocklin, M. 2015, in Proceedings of the 14th Python in Science Conference, ed.
  K.~Huff \& J.~Bergstra, 130 -- 136

\bibitem[{Scherer {et~al.}(2022)Scherer, Cuadra, \& Bauer}]{Scherer2022}
Scherer, A., Cuadra, J., \& Bauer, F.~E. 2022, \aap, 659, A105,
  \dodoi{10.1051/0004-6361/202142401}

\bibitem[{{Scherer} {et~al.}(2023){Scherer}, {Cuadra}, \&
  {Bauer}}]{Scherer2023}
{Scherer}, A., {Cuadra}, J., \& {Bauer}, F.~E. 2023, \aap, 679, A114,
  \dodoi{10.1051/0004-6361/202245822}

\bibitem[{{Shalchi}(2021)}]{Shalchi-2021}
{Shalchi}, A. 2021, \apj, 923, 209, \dodoi{10.3847/1538-4357/ac2363}

\bibitem[{{Sofue}(2000)}]{Sofue00}
{Sofue}, Y. 2000, \apj, 540, 224, \dodoi{10.1086/309297}

\bibitem[{{The CTA Consortium}(2019)}]{CTA_Science}
{The CTA Consortium}. 2019, Science with the Cherenkov Telescope Array (WORLD
  SCIENTIFIC), \dodoi{10.1142/10986}

\bibitem[{Virtanen {et~al.}(2020)Virtanen, Gommers, Oliphant, Haberland, Reddy,
  Cournapeau, Burovski, Peterson, Weckesser, Bright, {van der Walt}, Brett,
  Wilson, Millman, Mayorov, Nelson, Jones, Kern, Larson, Carey, Polat, Feng,
  Moore, {VanderPlas}, Laxalde, Perktold, Cimrman, Henriksen, Quintero, Harris,
  Archibald, Ribeiro, Pedregosa, {van Mulbregt}, \& {SciPy 1.0
  Contributors}}]{2020SciPy-NMeth}
Virtanen, P., Gommers, R., Oliphant, T.~E., {et~al.} 2020, Nature Methods, 17,
  261, \dodoi{10.1038/s41592-019-0686-2}

\bibitem[{{W}es {M}c{K}inney(2010)}]{pandas}
{W}es {M}c{K}inney. 2010, in {P}roceedings of the 9th {P}ython in {S}cience
  {C}onference, ed. {S}t\'efan van~der {W}alt \& {J}arrod {M}illman, 56 -- 61,
  \dodoi{10.25080/Majora-92bf1922-00a}

\end{thebibliography}
